\documentclass[a4paper,onecolumn,11pt]{quantumarticle}
\pdfoutput=1

\usepackage{tikzit}

\tikzstyle{diamant}=[diamond, fill=couleurdefond, draw=black]
\tikzstyle{newe}=[rectangle, fill={gray!15}, draw=black, tikzit shape=rectangle, inner sep=0.2em]
\tikzstyle{cercle}=[circle, fill=couleurdefond, draw=black]
\tikzstyle{scercle}=[circle, fill=couleurdefond, draw=black, tikzit fill=white, inner sep=0.1em]
\tikzstyle{cartouche}=[rounded rectangle, fill=couleurdefond, draw=black]
\tikzstyle{neg}=[rounded rectangle, fill=couleurdefond, draw=black, execute at end node={$\neg$}]
\tikzstyle{sneg}=[rounded rectangle, fill=couleurdefond, draw=black, execute at end node={$\neg$}, scale=0.8]
\tikzstyle{negserie}=[rounded rectangle, fill=couleurdefond, draw=black, execute at end node={\footnotesize$\star\star$}]
\tikzstyle{diagrammevide}=[rectangle, fill=couleurdefond, draw=black, inner sep=1.25em, borddiagrammevide, tikzit shape=rectangle]
\tikzstyle{mdiagrammevide}=[rectangle, fill=couleurdefond, draw=black, inner sep=0.75em, sborddiagrammevide, tikzit shape=rectangle]
\tikzstyle{msdiagrammevide}=[rectangle, fill=couleurdefond, draw=black, inner sep=0.7em, msborddiagrammevide, tikzit shape=rectangle]
\tikzstyle{sdiagrammevide}=[rectangle, fill=couleurdefond, draw=black, inner sep=0.5em, sborddiagrammevide, tikzit shape=rectangle]
\tikzstyle{xsdiagrammevide}=[rectangle, fill=couleurdefond, draw=black, inner sep=0.4em, xsborddiagrammevide, tikzit shape=rectangle]
\tikzstyle{bs}=[shape=beam, fill=couleurdefond, draw, inner sep=0.25em, thick, tikzit fill=white]
\tikzstyle{sbs}=[shape=beam, fill=couleurdefond, draw, inner sep=0.2em, thick, tikzit fill=white]
\tikzstyle{npbs}=[shape=beam, horizontal fill={{npbsmoitiebasse}{npbsmoitiehaute}}, draw, inner sep=0.25em, thick, tikzit fill={rgb,255: red,128; green,128; blue,128}]
\tikzstyle{npbsalenvers}=[shape=beam, horizontal fill={{npbsmoitiehaute}{npbsmoitiebasse}}, draw, inner sep=0.25em, thick, tikzit fill={rgb,255: red,128; green,128; blue,128}]
\tikzstyle{snpbs}=[shape=beam, horizontal fill={{npbsmoitiebasse}{npbsmoitiehaute}}, draw, inner sep=0.2em, thick, tikzit fill={rgb,255: red,128; green,128; blue,128}]
\tikzstyle{snpbsalenvers}=[shape=beam, horizontal fill={{npbsmoitiehaute}{npbsmoitiebasse}}, draw, inner sep=0.2em, thick, tikzit fill={rgb,255: red,128; green,128; blue,128}]
\tikzstyle{cnot}=[shape=circle, draw, path picture={ 
\draw[black](path picture bounding box.north) -- (path picture bounding box.south) (path picture bounding box.west) -- (path picture bounding box.east);
}, tikzit fill={rgb,255: red,223; green,223; blue,223}]
\tikzstyle{thickcnot}=[shape=circle, draw, thick, path picture={ 
\draw[thick,black](path picture bounding box.north) -- (path picture bounding box.south) (path picture bounding box.west) -- (path picture bounding box.east);
}, tikzit fill={rgb,255: red,223; green,223; blue,223}]
\tikzstyle{boite22}=[fill=white, draw=black, shape=rectangle, minimum height=1cm, minimum width=0.5cm]
\tikzstyle{boite15}=[fill=white, draw=black, shape=rectangle, minimum height=0.7cm, minimum width=0.5cm]
\tikzstyle{boite2}=[fill=white, draw=black, shape=rectangle, minimum height=0cm, minimum width=0cm]
\tikzstyle{snegpotentiel}=[fill=couleurdefond, draw=black, shape=rounded rectangle, inner sep=0.25em, tikzit fill={rgb,255: red,191; green,191; blue,191}, execute at end node={\footnotesize$\star$}]
\tikzstyle{negpotentiel}=[fill=couleurdefond, draw=black, shape=rounded rectangle, tikzit fill={rgb,255: red,191; green,191; blue,191}, execute at end node={$\star$}]
\tikzstyle{token}=[fill=black, draw=black, shape=circle, inner sep=0.1em]
\tikzstyle{whitetoken}=[fill=white, draw=black, shape=circle, inner sep=0.1em]
\tikzstyle{boitePBS}=[fill=white, draw=gray, thick, shape=rectangle, rounded corners=3pt, minimum height=0.6cm, inner sep=0.1em, minimum width=0.5cm]
\tikzstyle{boitePBS2}=[fill=white, draw=gray, thick, shape=rectangle, rounded corners=3pt, minimum height=0.55cm, inner sep=0.1em, minimum width=0.5cm]
\tikzstyle{sgene}=[fill={gray!30}, draw=black, shape=rounded rectangle, rounded rectangle east arc=0pt, minimum height=0.5cm, inner sep=0em, minimum width=0cm, scale=0.8]
\tikzstyle{sdetector}=[fill={gray!30}, draw=black, shape=rounded rectangle, rounded rectangle west arc=0pt, minimum height=0.5cm, inner sep=0em, minimum width=0cm, scale=0.8]
\tikzstyle{xsgene}=[fill={gray!30}, draw=black, shape=rounded rectangle, rounded rectangle east arc=0pt, minimum height=0.5cm, inner sep=0em, minimum width=0cm, scale=0.67]
\tikzstyle{xsdetector}=[fill={gray!30}, draw=black, shape=rounded rectangle, rounded rectangle west arc=0pt, minimum height=0.5cm, inner sep=0em, minimum width=0cm, scale=0.67]
\tikzstyle{PolRot}=[fill={gray!30}, draw=black, shape=rectangle, minimum height=0.5cm, inner sep=0.1em, minimum width=0.1cm]
\tikzstyle{PhS}=[fill=white, draw=black, shape=rectangle, minimum height=0.5cm, inner sep=0.1em, minimum width=0.1cm]
\tikzstyle{gene}=[fill={gray!30}, draw=black, shape=rounded rectangle, rounded rectangle east arc=0pt, minimum height=0.5cm, inner sep=0em, minimum width=0cm]
\tikzstyle{detector}=[fill={gray!30}, draw=black, shape=rounded rectangle, rounded rectangle west arc=0pt, minimum height=0.5cm, inner sep=0em, minimum width=0cm]
\tikzstyle{cartoucherouge}=[rounded rectangle, fill={red!55!white}, draw=black, tikzit fill=red]
\tikzstyle{cartouchebleu}=[rounded rectangle, fill={blue!33!white}, draw=black, tikzit fill=blue]
\tikzstyle{diamantrouge}=[diamond, fill={rgb,255: red,255; green,115; blue,115}, draw=black]
\tikzstyle{diamantbleu}=[diamond, fill={rgb,255: red,171; green,171; blue,255}, draw=black]
\tikzstyle{control}=[fill=black, draw=black, shape=circle, scale=0.35]
\tikzstyle{wcontrol}=[fill=white, draw=black, shape=circle, scale=0.35]

\tikzstyle{new}=[-, tikzit draw=magenta]
\tikzstyle{tirets}=[-, draw=black, dashed]
\tikzstyle{noire}=[-, draw=black, tikzit draw=magenta]
\tikzstyle{ep}=[-, draw=black, tikzit draw=magenta]
\tikzstyle{longdashed}=[-, dash pattern=on 5pt off 5pt]
\tikzstyle{pointilles}=[-, draw=black, dotted]
\tikzstyle{grise}=[-, draw={rgb,255: red,191; green,191; blue,191}]
\tikzstyle{rouge}=[-, draw=red]
\tikzstyle{bleue}=[-, draw=bleu, tikzit draw=blue]
\tikzstyle{verte}=[-, draw={rgb,255: red,0; green,230; blue,0}]
\tikzstyle{borddiagrammevide}=[-, dash pattern=on 0.5em off 0.5em on 0.5em off 0.5em on 0.5em off 0em]
\tikzstyle{msborddiagrammevide}=[-, dash pattern=on 0.28em off 0.28em on 0.28em off 0.28em on 0.28em off 0em]
\tikzstyle{sborddiagrammevide}=[-, dash pattern=on 0.2em off 0.2em on 0.2em off 0.2em on 0.2em off 0em]
\tikzstyle{xsborddiagrammevide}=[-, dash pattern=on 0.1em off 0.1em on 0.15em off 0.1em on 0.1em off 0em]
\tikzstyle{mediumdash}=[-, dash pattern=on 2pt off 2pt]
\tikzstyle{rougefonce}=[-, draw={red!50!black}, tikzit draw={rgb,255: red,136; green,0; blue,0}]

\input{styles-pbs.tikzdefs}
\input{env.sty}

\usepackage[figuresright]{rotating}

\begin{document}

\title{Strong Simulation of Linear Optical Processes}

\author{Nicolas Heurtel}
\email{nicolas.heurtel@quandela.com}
\affiliation{Quandela, 7 Rue Léonard de Vinci, 91300 Massy, France}
\affiliation{Université Paris-Saclay, Inria, CNRS, ENS Paris-Saclay, CentraleSupélec, LMF, 91190, 15 Gif-sur-Yvette, France}

\author{Shane Mansfield}%
\email{shane.mansfield@quandela.com}

\author{Jean Senellart}
\email{jean.senellart@quandela.com}
 
\affiliation{Quandela, 7 Rue Léonard de Vinci, 91300 Massy, France}%

\author{Benoît Valiron}
\email{benoit.valiron@centralesupelec.fr}
\affiliation{Université Paris-Saclay, Inria, CNRS, ENS Paris-Saclay, CentraleSupélec, LMF, 91190, 15 Gif-sur-Yvette, France}

\maketitle

\begin{abstract}
  \noindent In this paper, we provide an algorithm and general framework
  for the simulation of photons passing through linear optical interferometers.
  Given $n$ photons at the input of an $m$-mode interferometer, our algorithm
  computes the probabilities of all possible output states with time complexity
  \bigO{n\CMn}, linear in the number of output states \CMn. It outperforms
  the \naive method by an exponential factor, and for the restricted problem of
  computing the probability for one given output it improves the time complexity over the state-of-the-art for the permanent of matrices with multiple rows or columns, with a tradeoff in the memory usage. Our algorithm also has additional versatility by virtue of its
  use of memorisation -- the storing of intermediate results -- which is
  advantageous in situations where several input states may be of interest.
  Additionally it allows for hybrid simulations, in which outputs are sampled
  from output states whose probability exceeds a given threshold, or from a
  restricted set of states. We consider a concrete, optimised implementation, and we benchmark the efficiency of our approach compared to existing tools.
\end{abstract}



\section{\label{sec:level1} Introduction}

In quantum computation one encodes information into the states of quantum
systems -- photons, atoms, ions, etc.\ --
which can then processed by evolving and manipulating those
systems according to the laws of quantum mechanics.
It is by now well-known that the paradigm opens vast possibilities for exploiting non-classical behaviours
available to quantum systems in order to process information in radically new
ways that can lead to a variety of \emph{quantum advantages} including computational speedups \cite{shor1994algorithms}, enhanced
security \cite{bennet1984quantum}, more efficient communication \cite{brassard2003quantum}, and the potential for reduced energy consumption \cite{auffeves2022quantum,jaschke2022quantum}, when compared to classical
information processing.

The development quantum technologies aiming to leverage such advantages has been advancing at pace over the past
number of years.
A variety of different hardwares, each using different physical supports for the
quantum information, are being pursued.
Among these, photonic hardware has a privileged role in the sense that regardless of
hardware choice it will eventually be necessary to network quantum processors,
and as the only viable support for communicating quantum information it is
inevitable that some quantum information must eventually be treated photonically.
Photons have a number of other desirable features too,
including an absence of decoherence in transparent media
-- i.e.\ a capacity to reliably maintain their quantum states --, reduced cryogenic requirements compared to other hardware approaches, and good
prospects for scalability due to compatibility with the existing semiconductor
industry \cite{optimistic_Rudolph}.

Photonic quantum technologies consisting of single-photon sources, which are coupled to
linear optical interferometers -- which may be parametrisable and take the form
of integrated circuits --, which are coupled in turn to photon detectors, offer a promising route to implementing quantum computation.
They enable both non-universal models of quantum computation
\cite{aaronson_computational_2011}, which have led to laboratory demonstrations that claim
to show quantum
computational advantages with today's technology \cite{zhong_quantum_2020,wu_strong_2021},
as well as models for achieving universal
\cite{klm_scheme_2001}, and
fault-tolerant quantum computation \cite{bartolucci_fusion-based_2021}.

To accompany the technological developments in photonic quantum computing it is
also important to have access to tools for design, testing, and experimenting
with algorithms, protocols, and schemes.
In this respect, the classical simulation of photonic quantum computing
platforms has become an increasingly important problem.
Of course, one of the main interests of quantum computation is that it quickly
becomes unfeasible for classical processors to simulate.
Yet there are clear benefits to achieving optimal classical simulation
within the theoretical limits.
This can aid in designing and perfecting interferometers that
generate specific logic gates, entangled states, and other building-block
components of quantum computers. It can provide both development and verification
tools for algorithm and software development. Furthermore it can help to
define the performance boundaries separating the quantum from the classical
computational paradigm.

Two special cases are of particular interest: strong simulation, where a
classical program
computes the complete quantum state obtained at the outputs of a photonic
circuit;
and weak simulation, or sampling, where the classical program emulates the
probabilistic behaviour that would be observed at the outputs.
The task of predicting, with classical algorithms and computers, the output
states for photons passing through interferometers relates to the $\# P$-hard
problem of calculating the permanents of complex matrices
\cite{valiant1979complexity} associated with the interferometer
\cite{Scheel2004PermanentsIL}.

  In order to perform strong simulation, one solution is 
  therefore to reuse algorithms originally designed for computing
  permanents.
The state-of-the-art classical algorithms for computation of the permanent of a complex
matrix are those due to Ryser \cite{ryser1963combinatorial} and Glynn
\cite{glynn2010permanent}, while the state-of-the-art classical algorithms for boson sampling (essentially the weak simulation problem) are those due to
Clifford and Clifford \cite{clifford_classical_2018,clifford2020faster}. As the strong simulation of linear optical circuits involves the permanent computation of matrices with repeated rows or columns, algorithms with a better complexity \cite{Shche2019} than Ryser's or Glynn's can be used.

  On the other hand and independently from permanents, the folklore of
  linear optics is aware of informal, back-of-the-envelope techniques
  for running fixed-photon number simulations. The goal of this paper
  is to formalize and analyze these folklore techniques and provide a
  comparison with the algorithms for the permanents of complex
  matrices.
In particular, we propose a framework for strong simulation, amenable to weak
simulation, which additionally allow for hybrid forms of simulation between
these two particular cases. Accross the paper, we shall be using the acronym SLOS, standing for \emph{Strong Linear Optical Simulator}.
We derive an algorithm for computing the permanent of a
complex matrix with repeated rows that improves the state-of-the-art \cite{Shche2019} time complexity, with a tradeoff in memory.
We detail optimised implementations of our algorithms that can be found in the \emph{QuandeLibC} library\footnote{On Github at {\url{https://github.com/Quandela/QuandeLibC}}.} (which is also integrated in the open-source software platform \emph{Perceval} \cite{heurtel_perceval_2022}).
We benchmark the performance of our algorithms,
comparing them to implementations of the Glynn and Ryser algorithms in the
\emph{the Walrus} library \cite{gupt2019walrus}, as well as implementations in \emph{QuandeLibC} of these
algorithms, which appear to be more efficient, and of the Clifford and Clifford algorithms also in \emph{QuandeLibC}.

\paragraph{Contributions}
A preliminary version of this work has been presented at IEEE Quantum Computing and Engineering \cite{heurtel2022sloi}. The main contributions of this paper are three Strong Linear Optical Simulators (\SLOS), and can be summarised as follows:

\begin{itemize}
    \item An algorithm, labelled \SLOSfull, for computing the full output amplitude distribution of a linear optical circuit for a given input. Although this method would be naturally described in physics textbooks, to the authors' knowledge this is the first complexity study and explicit implementation of that method. The time complexity is $\bigO{n\CMn}$, and so is linear in the number of output states \CMn. The full distribution can be obtained in an optimal space of $\mathcal{O}\CMn$. 
    \item A generalised strong simulation algorithm, labelled \SLOSgen, for computing the amplitudes of any set of outputs from any set of inputs. For one input, \SLOSgen has the same time complexity as \SLOSfull for the full output distribution, and improves the state-of-the-art \cite{Shche2019} for the specific case of a single output, giving a new upper bound for computing a permanent with repeated rows or columns.
    \item A hybrid algorithm that can combine both weak and strong simulation for many inputs and outputs, labelled \SLOShyb. \SLOShyb\ can sample from a set of outputs whose probability exceeds a given threshold, or even sample among a restricted set of states. \SLOShyb\ can both perform both weak sampling and strong simulation (as \SLOSgen).
    \item Detailed optimised implementations of both \SLOSfull and \SLOSgen. The implementations are open-source and available in the {\it QuandeLibC} library.
    \item Practical performance benchmarking of the  algorithm  in  a  generic  example  of  a  quantum  machine learning  algorithm \cite{gan2021fock},  where  it  is  seen  to  give  a  considerable practical edge over the \naive approach to simulation.
\end{itemize}

\paragraph{Plan}
The paper is structured as follows: in \cref{sec:level2} we provide some background on linear optical simulation, and in particular the specific problems we focused on are set up in \cref{sec2:SLOSproblems}, with illustrations of typical use cases. 
In \cref{sec3} we introduce the algorithms and their complexity analysis, summarised in Table \ref{tab:complexity}. 
The practical implementation and optimisation are presented in \cref{sec:implem}, while benchmarks with the permanent-based model are in \cref{sec:performance}.
We finally conclude and discuss in \cref{sec:conclusion}.



\section{\label{sec:level2} Simulating Linear Optical Circuits}

In \cref{sec2:notations}, we set up some formalism and notational conventions to be used throughout the paper.
After briefly explaining the hardness of LO-circuit simulation in \cref{sec2:hardness}, we present and define the weak and strong simulation problems in \cref{sec2:weakandstrong}.
In \cref{sec2:SLOSproblems}, we present the two strong simulation problems that we propose to answer in \cref{sec3}.

\subsection{Formalism of Linear Optical Circuits and Notation}
\label{sec2:notations}

Throughout this paper, we will be considering $n$ indistinguishable photons over $m$ modes. Typically these are spatial modes, but they could in principle also correspond to other discrete degrees of freedom such as polarisation, frequency, or time-bins \cite{KMN+2007Linear,KL2010Introduction}. States of the system will be Fock states or their superpositions and we write \lket{s}{}{}{m} to denote the Fock state with $s_i$ photons in mode $i$. Sometimes it will be interesting to consider states containing less than $n$ photons, so we introduce the notation \lket{s}{}{k}{m} to describe a state with $\sum_{i=1}^{m}s_i=k$ photons, sometimes shortening this to \sket{s}{m}{k}. The vacuum state, with no photon in $m$ modes, will be denoted as $\sket{0}{m}{}$. We also introduce $\F{m}{k}$ as the set of the Fock states of $k$ photons into $m$ modes, so that $\F{m}{k}= \left\{ \lket{s}{}{k}{m}  \middle| s_i \in \mathbb{N} \right\}$. It is known that $\#\F{m}{k}=\binom{k+m-1}{m-1}$, as it is exactly the number of ways to put $k$ indistinguishable balls into $m$ distinguishable bins \cite{feller1}. For readability and as $m$ will be a fixed parameter, that number will denoted as \Mk.

It is standard in the second quantisation formalism to associate each mode $i$ with a creation operator $\opc{a}{i}: \F{m}{k} \to \F{m}{k+1}$ 
acting as follows:

$$\opc{a}{i} \ket{s_1,\dots,s_i,\dots,s_m}^k=\sqrt{s_i+1}\ket{s_1,\dots,s_i+1,\dots,s_m}^{k+1}$$

This paper focuses only on linear optical operations, for which the transformations on the creation operators are described by a unitary matrix $U=(u_{i,j})$ of size $m \times m$ such that $\opc{a}{p} \mapsto \sum_{\added{i}=1}^m u_{i,p} \opc{a}{i}$.
As shown in \cite{Reck94}, every such unitary can be implemented by a linear optical circuit of $m$ spatial modes, with only phase shifters and beamsplitters \cite{KMN+2007Linear} as linear optical components. We will call such a circuit an LO-circuit.

Each unitary matrix $\U{}:m\times m$ acting on the vector of creation operators can be associated with a unitary operator \Ua{} on $\mathcal{H}_m=\bigoplus_{k=0}^{+\infty}\mathcal{H}_m^k$ where $\mathcal{H}_m^k$ is the Hilbert space generated by the elements of \F{m}{k} \cite{aaronson_computational_2011,Scheel2004PermanentsIL}. 
$\Ua\sket{s}{m}{k}$ will represent the state obtained when the state \sket{s}{m}{k} is the input of an LO-circuit implementing $U$ and can be obtained with: 
\begin{equation} \label{eq:output}
    \Ua \sket{s}{m}{n} =  \prod_{p=1}^{m} \frac{1}{\sqrt{s_p!}}\left(\sum_{i=1}^{m} u_{i,p}\opc{a}{i}\right)^{s_p} \sket{0}{m}{}
\end{equation}

Following our notation, $\sbra{t}{m}{n} \Ua \sket{s}{m}{n}$ is the amplitude assigned to the state $\sket{t}{m}{n}$ within the overall output state \Ua\sket{s}{m}{n}.

\subsection{Hardness of Linear Optical Simulations}
\label{sec2:hardness}

Given a matrix $M:n\times n$, the permanent of $M$ is defined as follows: 
$$\mathrm{Perm}(M) = \sum_{\sigma \in S_n}\prod_{i=1}^n m_{i,\sigma(i)}.$$
Efficiently computing the permanent is crucial in evaluating linear optical transformations, as we can show that:

\begin{equation}{\label{eq:permcoef}}
    \sbra{t}{m}{n} \Ua \sket{s}{m}{n} = \frac{\mathrm{Perm}\left(U_{\ket{s},\ket{t}}\right)}{\sqrt{s_1! \dots s_m!t_1!\dots t_m!}}.
\end{equation}
where $U_{\ket{s},\ket{t}}$ is obtained from the unitary $U$ by repeating $s_{i}$ times its $i^{\text{th}}$ column and $t_{j}$ times its $j^{\text{th}}$ row \cite{osti_4420279,Aaronson_2011}. We return to this in  \cref{sec2:SLOSproblems}.

It is known that computing a permanent of a general complex matrix \cite{valiant1979complexity} or even subclasses of real orthogonal matrices \cite{GrierS18} is a $\#P$-hard problem. The fastest known algorithms \cite{ryser1963combinatorial,glynn2010permanent} for computing a permanent of a general matrix of size $n$, with some pre-computation allowed \cite{Nijenhuis1978CombinatorialAF}, are in $O(n2^{n})$. 

For computing the permanent in \cref{eq:permcoef}, the redundancy of the rows and columns in $U_{\ket{s},\ket{t}}$ can be taken advantage of, giving a faster algorithm \cite{Shche2019}. More generally, for an output state $\ket{t_1,...,t_m}$, the permanent can be computed in:
\begin{equation}
    \label{eq:permanentmin}
    \bigO{\minperm}
\end{equation}

operations. Note that this bound is equal to $\bigO{n2^{n-1}}$ in the worst case, when there is at most one photon per mode, and equal to $\bigO{n}$ in the best case, when only one mode is occupied.

\subsection{Weak and Strong Simulation}
\label{sec2:weakandstrong}

When considering a model for a biased coin, two strategies can be followed. Either one can try to literally emulate the probabilistic behavior of the coin and have a protocol answering ``head'' or ``tail'' with the appropriate probabilities, or to more fully characterise the behaviour and by listing the (two) precise probabilities for ``head'' and ``tail''.
The former approach is called \emph{weak simulation} while the latter is \emph{strong simulation}.

\paragraph{Weak Simulation}
Weak simulation of LO-circuits is the classical sampling from their output distribution, also known as the Boson Sampling problem \cite{aaronson_computational_2011}. Given an input \sket{s}{m}{n} and a LO-circuit implementing a unitary \U, we would like to sample an output $\sket{t}{m}{n}$ from the distribution $ \mathcal{D}_U(s) = \left \{ \left\lvert \sbra{t}{}{} \Ua \sket{s}{}{} \right\rvert^2, \sket{t}{}{} \in \F{m}{n}  \right \}  $. 

Under some assumptions, weak simulation has been shown to be classically hard \cite{aaronson_computational_2011}, as it would imply $P^{\#P}=BPP^{NP}$ leading to a collapse of the polynomial hierarchy of complexity classes to the third level. Therefore, Boson Sampling is a good candidate for quantum advantage, as a linear optical computer with single-photon inputs can naturally sample from $\mathcal{D}_U(s)$.
It was shown that weak simulation could be done in $O(n2^n + mn^2)$ \cite{clifford_classical_2018}, which was further improved to  $O(n1.69^n)$ on average when $m=n$ \cite{clifford2020faster} thanks to a more efficient way to compute permanents with repeated rows.

Even though weak simulation is not the main focus of the paper, it can be recovered as a by-product of the general algorithm \SLOShyb presented in \cref{subsec:SLOShyb}.

\paragraph{Strong Simulation}
Strong simulation of LO-circuits is the classical computation of the output amplitudes (or the probabilities): Given an input \sket{s}{m}{n} and an LO-circuit implementing a unitary \U, we would like to compute the amplitudes \tUs , or the probabilities \tUsprob with $\ket{t} \in \F{m}{n}$.

As explained in \cref{sec2:hardness}, we can directly compute them by computing the permanents of $U_{\ket{s},\ket{t}}$. Therefore, the complexity of computing one amplitude or probability is exactly the complexity of computing one permanent.

In this paper, we propose a procedure directly computing the amplitudes of several outputs or inputs, which is more efficient than computing them separately and independently, cf \cref{subsec:SLOSgen,subsec:SLOSfull} , only needing $\bigO{\sum_{i=1}^mt_i\prod_{j\neq i}(t_j+1)}$ operations for one output, as shown in \cref{subsec:oneoutonein} and summarised in Table \ref{tab:complexity}.
We therefore can efficiently solve two kinds of strong simulation problems, which we set out in \cref{sec2:SLOSproblems}.

\subsection{\SLOS Problems: Two Strong Linear Optical Simulation Problems}
\label{sec2:SLOSproblems}

\begin{figure}
     \centering
     \begin{subfigure}[b]{0.3\textwidth}
         \centering
         \scalebox{0.7}{\begin{tikzpicture}
	\begin{pgfonlayer}{nodelayer}
		\node [style=none] (0) at (-2.75, 2.75) {};
		\node [style=none] (1) at (-2.75, -2.75) {};
		\node [style=none] (2) at (2.75, 2.75) {};
		\node [style=none] (3) at (2.75, -2.75) {};
		\node [style=none] (4) at (0, 0) {$LO-circuit$};
		\node [style=none] (5) at (-3.75, 2) {};
		\node [style=none] (6) at (-2.75, 2) {};
		\node [style=none] (7) at (2.75, 2) {};
		\node [style=none] (8) at (3.75, 2) {};
		\node [style=none] (9) at (-3.75, 1) {};
		\node [style=none] (10) at (-2.75, 1) {};
		\node [style=none] (11) at (-3.75, 0) {};
		\node [style=none] (12) at (-2.75, 0) {};
		\node [style=none] (13) at (-3.75, -1) {};
		\node [style=none] (14) at (-2.75, -1) {};
		\node [style=none] (15) at (2.75, 1) {};
		\node [style=none] (16) at (3.75, 1) {};
		\node [style=none] (17) at (2.75, 0) {};
		\node [style=none] (18) at (3.75, 0) {};
		\node [style=none] (19) at (2.75, -1) {};
		\node [style=none] (20) at (3.75, -1) {};
		\node [style=none] (25) at (-3.75, -2) {};
		\node [style=none] (26) at (-2.75, -2) {};
		\node [style=none] (31) at (2.75, -2) {};
		\node [style=none] (32) at (3.75, -2) {};
		\node [style=none] (41) at (4.5, 2) {*};
		\node [style=none] (42) at (4.5, 1) {*};
		\node [style=none] (43) at (4.5, 0) {*};
		\node [style=none] (44) at (4.5, -1) {*};
		\node [style=none] (45) at (4.5, -2) {*};
		\node [style=none] (47) at (-4.75, 2) {$\bullet$};
		\node [style=none] (48) at (-4.75, 1) {$\bullet$};
		\node [style=none] (49) at (-4.75, 0) {$\bullet$};
		\node [style=none] (50) at (-4.75, -1) {$\bullet$};
		\node [style=none] (51) at (-4.75, -2) {$\bullet$};
		\node [style=none] (53) at (-4.75, 3.25) {$\overbrace{}$};
		\node [style=none] (54) at (-4.75, 4) {input};
		\node [style=none] (55) at (4.5, 3.25) {$\overbrace{}$};
		\node [style=none] (56) at (4.5, 4) {outputs};
	\end{pgfonlayer}
	\begin{pgfonlayer}{edgelayer}
		\draw (0.center) to (2.center);
		\draw (2.center) to (3.center);
		\draw (3.center) to (1.center);
		\draw (1.center) to (0.center);
		\draw (5.center) to (6.center);
		\draw (7.center) to (8.center);
		\draw (9.center) to (10.center);
		\draw (11.center) to (12.center);
		\draw (13.center) to (14.center);
		\draw (15.center) to (16.center);
		\draw (17.center) to (18.center);
		\draw (19.center) to (20.center);
		\draw (25.center) to (26.center);
		\draw (31.center) to (32.center);
	\end{pgfonlayer}
\end{tikzpicture}}
         \caption{Full distribution - one input}
         \label{figSLOSprobs1}
     \end{subfigure}
     \begin{subfigure}[b]{0.3\textwidth}
         \centering
         \scalebox{0.7}{\begin{tikzpicture}
	\begin{pgfonlayer}{nodelayer}
		\node [style=none] (0) at (-2.75, 2.75) {};
		\node [style=none] (1) at (-2.75, -2.75) {};
		\node [style=none] (2) at (2.75, 2.75) {};
		\node [style=none] (3) at (2.75, -2.75) {};
		\node [style=none] (4) at (0, 0) {$LO-circuit$};
		\node [style=none] (5) at (-3.75, 2) {};
		\node [style=none] (6) at (-2.75, 2) {};
		\node [style=none] (7) at (2.75, 2) {};
		\node [style=none] (8) at (3.75, 2) {};
		\node [style=none] (9) at (-3.75, 1) {};
		\node [style=none] (10) at (-2.75, 1) {};
		\node [style=none] (11) at (-3.75, 0) {};
		\node [style=none] (12) at (-2.75, 0) {};
		\node [style=none] (13) at (-3.75, -1) {};
		\node [style=none] (14) at (-2.75, -1) {};
		\node [style=none] (15) at (2.75, 1) {};
		\node [style=none] (16) at (3.75, 1) {};
		\node [style=none] (17) at (2.75, 0) {};
		\node [style=none] (18) at (3.75, 0) {};
		\node [style=none] (19) at (2.75, -1) {};
		\node [style=none] (20) at (3.75, -1) {};
		\node [style=none] (21) at (-3.75, -2) {};
		\node [style=none] (22) at (-2.75, -2) {};
		\node [style=none] (23) at (2.75, -2) {};
		\node [style=none] (24) at (3.75, -2) {};
		\node [style=none] (25) at (4.5, 2) {*};
		\node [style=none] (26) at (4.5, 1) {*};
		\node [style=none] (27) at (4.5, 0) {*};
		\node [style=none] (30) at (-4.75, 2) {$\bullet$};
		\node [style=none] (31) at (-4.75, 1) {$\bullet$};
		\node [style=none] (32) at (-4.75, 0) {$\bullet$};
		\node [style=none] (33) at (-4.75, -1) {$\bullet$};
		\node [style=none] (34) at (-4.75, -2) {$\bullet$};
		\node [style=none] (35) at (-4.75, 3.25) {$\overbrace{}$};
		\node [style=none] (36) at (-4.75, 4) {inputs};
		\node [style=none] (37) at (4.5, 3.25) {$\overbrace{}$};
		\node [style=none] (38) at (4.5, 4) {outputs};
		\node [style=none] (39) at (4.5, -1) {$\bullet$};
		\node [style=none] (40) at (4.5, -2) {$\bullet$};
	\end{pgfonlayer}
	\begin{pgfonlayer}{edgelayer}
		\draw (0.center) to (2.center);
		\draw (2.center) to (3.center);
		\draw (3.center) to (1.center);
		\draw (1.center) to (0.center);
		\draw (5.center) to (6.center);
		\draw (7.center) to (8.center);
		\draw (9.center) to (10.center);
		\draw (11.center) to (12.center);
		\draw (13.center) to (14.center);
		\draw (15.center) to (16.center);
		\draw (17.center) to (18.center);
		\draw (19.center) to (20.center);
		\draw (21.center) to (22.center);
		\draw (23.center) to (24.center);
	\end{pgfonlayer}
\end{tikzpicture}}
         \caption{General Heralded Scheme}
         \label{figSLOSprobs2}
     \end{subfigure}
     \begin{subfigure}[b]{0.3\textwidth}
         \centering
         \scalebox{0.7}{\begin{tikzpicture}
	\begin{pgfonlayer}{nodelayer}
		\node [style=none] (0) at (-2.75, 2.75) {};
		\node [style=none] (1) at (-2.75, -2.75) {};
		\node [style=none] (2) at (2.75, 2.75) {};
		\node [style=none] (3) at (2.75, -2.75) {};
		\node [style=none] (4) at (0, 0) {$LO-circuit$};
		\node [style=none] (5) at (-3.75, 2) {};
		\node [style=none] (6) at (-2.75, 2) {};
		\node [style=none] (7) at (2.75, 2) {};
		\node [style=none] (8) at (3.75, 2) {};
		\node [style=none] (9) at (-3.75, 1) {};
		\node [style=none] (10) at (-2.75, 1) {};
		\node [style=none] (11) at (-3.75, 0) {};
		\node [style=none] (12) at (-2.75, 0) {};
		\node [style=none] (13) at (-3.75, -1) {};
		\node [style=none] (14) at (-2.75, -1) {};
		\node [style=none] (15) at (2.75, 1) {};
		\node [style=none] (16) at (3.75, 1) {};
		\node [style=none] (17) at (2.75, 0) {};
		\node [style=none] (18) at (3.75, 0) {};
		\node [style=none] (19) at (2.75, -1) {};
		\node [style=none] (20) at (3.75, -1) {};
		\node [style=none] (21) at (-3.75, -2) {};
		\node [style=none] (22) at (-2.75, -2) {};
		\node [style=none] (23) at (2.75, -2) {};
		\node [style=none] (24) at (3.75, -2) {};
		\node [style=none] (28) at (-4.75, 2) {$\bullet$};
		\node [style=none] (29) at (-4.75, 1) {$\bullet$};
		\node [style=none] (30) at (-4.75, 0) {$\bullet$};
		\node [style=none] (31) at (-4.75, -1) {$\bullet$};
		\node [style=none] (32) at (-4.75, -2) {$\bullet$};
		\node [style=none] (33) at (-4.75, 3.25) {$\overbrace{}$};
		\node [style=none] (34) at (-4.75, 4) {inputs};
		\node [style=none] (35) at (4.5, 3.25) {$\overbrace{}$};
		\node [style=none] (36) at (4.5, 4) {outputs};
		\node [style=none] (39) at (4.5, 2) {$\bullet$};
		\node [style=none] (40) at (4.5, 1) {$\bullet$};
		\node [style=none] (41) at (4.5, 0) {$\bullet$};
		\node [style=none] (42) at (4.5, -1) {$\bullet$};
		\node [style=none] (43) at (4.5, -2) {$\bullet$};
	\end{pgfonlayer}
	\begin{pgfonlayer}{edgelayer}
		\draw (0.center) to (2.center);
		\draw (2.center) to (3.center);
		\draw (3.center) to (1.center);
		\draw (1.center) to (0.center);
		\draw (5.center) to (6.center);
		\draw (7.center) to (8.center);
		\draw (9.center) to (10.center);
		\draw (11.center) to (12.center);
		\draw (13.center) to (14.center);
		\draw (15.center) to (16.center);
		\draw (17.center) to (18.center);
		\draw (19.center) to (20.center);
		\draw (21.center) to (22.center);
		\draw (23.center) to (24.center);
	\end{pgfonlayer}
\end{tikzpicture}}
         \caption{General Post Selected Scheme}
         \label{figSLOSprobs3}
     \end{subfigure}
        \caption{Application Schemes of Problem 1 and Problem 2. $\bullet$ is a chosen number of photons, while $*$ is every possible number }
        \label{figSLOSprobs}
\end{figure}
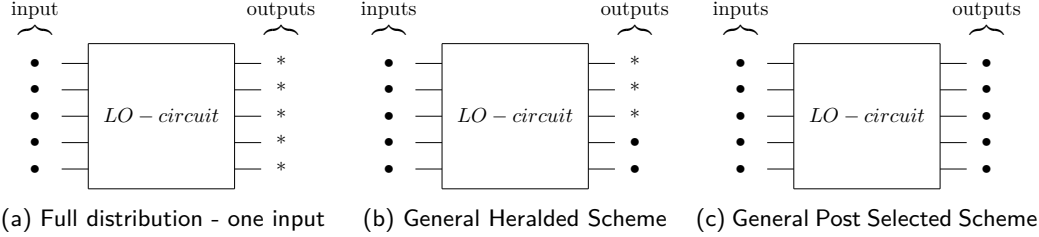

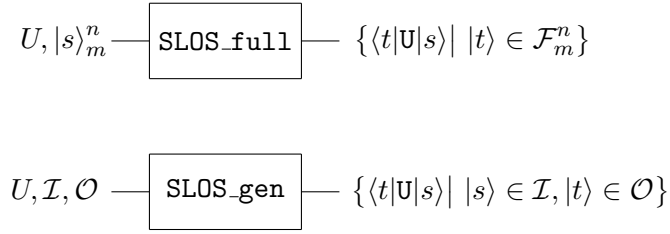
\begin{figure}

$$\begin{tikzpicture}
	\begin{pgfonlayer}{nodelayer}
		\node [style=none] (0) at (-2, 3) {};
		\node [style=none] (1) at (-2, 1) {};
		\node [style=none] (2) at (2, 3) {};
		\node [style=none] (3) at (2, 1) {};
		\node [style=none] (4) at (0, 2) {$\texttt{SLOS\_full}$};
		\node [style=none] (5) at (-3, 2) {};
		\node [style=none] (6) at (-2, 2) {};
		\node [style=none] (7) at (2, 2) {};
		\node [style=none] (8) at (3, 2) {};
		\node [style=none] (9) at (-4.25, 2) {$U, \ket{s}_m^n$};
		\node [style=none] (10) at (6.5, 2) {$ \displaystyle \big\{ \bra{t} \mathtt{U} \ket{s} \big| \ \ket{t} \in \mathcal{F}_m^n \big\}$ };
		\node [style=none] (11) at (-2, -1) {};
		\node [style=none] (12) at (-2, -3) {};
		\node [style=none] (13) at (2, -1) {};
		\node [style=none] (14) at (2, -3) {};
		\node [style=none] (15) at (0, -2) {$\texttt{SLOS\_gen}$};
		\node [style=none] (16) at (-3, -2) {};
		\node [style=none] (17) at (-2, -2) {};
		\node [style=none] (18) at (2, -2) {};
		\node [style=none] (19) at (3, -2) {};
		\node [style=none] (20) at (-4.5, -2) {$U, \mathcal{I},\mathcal{O}$};
		\node [style=none] (21) at (7.5, -2) {$ \big\{ \bra{t} \mathtt{U} \ket{s} \big| \  \ket{s} \in \mathcal{I}, \ket{t} \in \mathcal{O} \big\}$ };
	\end{pgfonlayer}
	\begin{pgfonlayer}{edgelayer}
		\draw (0.center) to (2.center);
		\draw (2.center) to (3.center);
		\draw (3.center) to (1.center);
		\draw (1.center) to (0.center);
		\draw (5.center) to (6.center);
		\draw (7.center) to (8.center);
		\draw (11.center) to (13.center);
		\draw (13.center) to (14.center);
		\draw (14.center) to (12.center);
		\draw (12.center) to (11.center);
		\draw (16.center) to (17.center);
		\draw (18.center) to (19.center);
	\end{pgfonlayer}
\end{tikzpicture}$$
\caption{Input and Output relations of \SLOSfull and \SLOSgen, solution algorithms respectively to the Problem 1 and Problem 2 of \cref{prob1} \label{fig:SLOSfamily}}
\end{figure}

This section will introduce the two classes LO simulation problems summarised in \cref{figSLOSprobs}. For each class, we provide concrete examples and typical use-cases.
Each problem has a proposed solution described in \cref{sec3} and summarised in \cref{fig:SLOSfamily}.

\paragraph{Problem 1: Full Amplitude List Simulation} \label{prob1}
Given an input \sket{s}{m}{n} and an LO-circuit of $m$ modes implementing a unitary matrix $\U:m\times m$, what is the output state \Ua \sket{s}{m}{n}? Equivalently, what is the full amplitude list $\left \{  \sbra{t}{}{} \Ua \sket{s}{}{}, \sket{t}{}{} \in \F{m}{n}  \right \}  $?
This situation is schematised in \cref{figSLOSprobs1}.

\paragraph{Example}
For instance, we can consider an LO-circuit of three modes implementing a unitary \U, with an input of two photons $\ket{1,1,0}$. In that problem, we would like to compute every output state of \F{3}{2}, meaning we would like to know the output:
\[\Ua\ket{1,1,0} = \alpha_1 \ket{2,0,0}+ \alpha_2\ket{0,2,0}+\alpha_3\ket{0,0,2}+\alpha_4 \ket{1,1,0}+\alpha_5 \ket{1,0,1}+\alpha_6 \ket{0,1,1}.\]
Notice that we can compute each amplitude $\alpha_i$ of an output state $\ket{t}$ by computing the permanent of $\U_{\ket{1,1,0},\ket{t}}$, as defined in \cref{sec2:hardness}. To compute $\alpha_3$, we would need to compute $\mathrm{Perm}(\U_{\ket{1,1,0},\ket{0,0,2}})$. By taking the first and second column of \U, $\begin{pmatrix}
u_{1,1} & u_{1,2}\\
u_{2,1} & u_{2,2} \\
u_{3,1} & u_{3,2}\\
\end{pmatrix}$and repeating two times the third row to form a $\U_{\ket{1,1,0},\ket{0,0,2}}$, we have:
$$\alpha_3=\frac{\mathrm{Perm}(\U_{\ket{1,1,0},\ket{0,0,2}})}{\sqrt{2!}}=\frac{\mathrm{Perm} \begin{pmatrix}
u_{3,1} & u_{3,2}\\
u_{3,1} & u_{3,2}\\
\end{pmatrix} }{\sqrt{2}} = \sqrt{2} \times u_{3,1}\times u_{3,2} $$

\paragraph{Typical Use-cases}
As the full amplitude list is a complete description of an LO-circuit and the input can be directly the state right after the photon sources, applications for this problem can be both practical and theoretical. One can think of: 

\begin{itemize}
\item Simulate or look for circuits preparing a specific output distribution like entangled high-dimensional states \cite{Gubarev20} or Bell Measurements \cite{Olivo_2018}.
\item Check the correctness and noise of experimental circuits with the theoretical distribution. One can certify the correctness of Boson sampling by computing metrics (such as total variational distance, bunching probabilities,\dots) from the statistics obtained \cite{GBSapprox21,Walschaers_2016,Shchesnovich_2021,Shchesnovich_2022}.
\item Machine Learning algorithms as the approximation of differential equations of \cite{gan2021fock}, application detailed in \cref{subsec:QML}.
\end{itemize}

\paragraph{Problem 2: Generic Strong Simulation}\label{prob2}
Given a set of inputs $\mathcal{I}$, a set of outputs $\mathcal{O}$ and an LO-circuit of $m$ modes implementing a unitary matrix $\U:m\times m$, what are the output amplitudes $\left \{  \sbra{t}{}{} \Ua \sket{s}{}{}, \sket{s}{}{} \in \mathcal{I}, \sket{t}{}{} \in \mathcal{O} \right \}  $?

\paragraph{Typical Use-cases}
Often, only specific outcomes are of interest, so the full output distribution of a LO-circuit is not needed. Also, for every circuit encoding logic gates or functions in general, we need to know the effect of the LO-circuit on each different possible input, so several inputs are needed. 
Therefore, that problem is very general and tackles various applications, in which we can highlight two general classes of schemes.

\begin{itemize}
\item Post-selected scheme. Presented in \cref{figSLOSprobs3}, it consists of only considering ---or selecting--- some inputs and some outputs according to some criteria. A canonical example is the implementation of the 2-qubit CNOT-gate of \cite{RalphCNOT}. Working on 6 modes with 2 photons, the considered input and output states are $X = \{\ket{0,1,0,1,0,0},\ket{0,1,0,0,1,0},\ket{0,0,1,1,0,0},\ket{0,0,1,0,1,0}\}$. In other words, the circuit is regarded as a map restricted on the 4-dimensional subspace $\mathbb{C}[X]$ generated by $X$, and not the full 21-dimensional $\mathcal{H}_6^2$ space. The term ``postselected'' refers to the fact that in general, the output state might have a component orthogonal to $\mathbb{C}[X]$ (for instance, maybe the state $\ket{0,2,0,0,0,0}$ has a non-zero amplitude). Such orthogonal components are considered bogus. They are ruled out at the end of the computation, when measuring the system. 
\item Heralded scheme. Presented in \cref{figSLOSprobs2}, it consists in designing a LO-circuit that can fail, but for which the failure can be decided upon the measurement result of some of the output modes. For instance, in \cite{KnillCZ} a scheme is proposed to implement the 2-qubit CZ-gate. The 4 input modes can accommodate 2, 3 or 4 photons, and the input state is generated with $\{\ket{1,1,1,1},\ket{1,0,1,1},\ket{0,1,1,1},\ket{0,0,1,1}\}$. For the output, we only consider the cases where the two last modes contain a photon: $\ket{*,*,1,1}$. This scheme is ``heralded'' in the sense that the computation only happen on the two first modes: the last two modes are just witnesses that the computation went well. Unlike the post-selected scheme, measuring these two last modes does not destruct possible entanglement on the two first modes.
\end{itemize}

\section{\SLOS Algorithms}
\label{sec3}
We first present \SLOSfull in \cref{subsec:SLOSfull}, which computes the full distribution of a given input, while also proving its complexity in time of $\bigO{n\Mn}$, where $M_n = \#\F{m}{n}=\binom{n+m-1}{m-1}$ as discussed in \cref{sec2:notations}. It is the setting of Problem~1 in \cref{sec2:SLOSproblems}.
We will then present \SLOSgen in \cref{subsec:SLOSgen}, when sets of inputs are outputs are allowed. It is the setting of Problem~2 in \cref{sec2:SLOSproblems}.
We conclude in \cref{subsec:SLOShyb} with the presentation of \SLOShyb, a simulation algorithm parameterised by a general cost function, and able to capture not only both weak and strong simulation, but also specific, crafted problems.

\subsection{ \SLOSfull: Computation of the Full Output Distribution of One Input}
\label{subsec:SLOSfull}

Given an input $ \sket{s}{m}{n}=\lket{s}{}{n}{m}$ and a unitary \U, $\Ua \sket{s}{m}{n}$ can be computed with \cref{eq:output}. As $\sum_{p=1}^ms_p=n$, the product only contains $n$ non-trivial terms, that can be arbitrary labeled as $p_1,p_2,\dots,p_n$, corresponding to the position of each photon. The key idea of \SLOSfull, presented in Algorithm~\ref{alg:slosfull}, is to decompose that product as follows:
\begin{equation} \label{eq:efficientoutput}
    \sum_{i_n=1}^m u_{i_n,p_n}\opc{a}{i}\left(\sum_{i_{n-1}=1}^m u_{i_{n-1},p_{n-1}}\opc{a}{i}\left(\dots\left(\sum_{i_1=1}^m u_{i_1,p_1}\opc{a}{i} \sket{0}{m}{} \right)\dots\right)\right)
\end{equation}
With this chosen order, we iteratively obtain the output of a state with $k+1$ photons from the output of a state with $k$ photons. To understand more closely how \SLOSfull works, let's first write the desired output state as: 
\[
\Ua \sket{s}{m}{n} = \Ua \left(\frac{ \opc{a}{p_n}\opc{a}{p_{n-1}}\dots \opc{a}{p_1}\sket{0}{m}{}}{ \prod_{p=1}^m\sqrt{s_p!}} \right) \]
and notice that for $k=0$ to $n-1$ we have:
\begin{equation} \label{eq:recstep}
\Ua \left(\opc{a}{p_{k+1}}\sket{s}{m}{k}\right)=  \sum_{i=1}^mu_{i,p_{k+1}} \opc{a}{i} \left(\Ua \sket{s}{m}{k}\right)
\end{equation}
Note that the normalisation factor can either be computed at each step $k$ (as in Algorithm~\ref{alg:slosfull}), or at the end for the final distribution $\Ua\sket{s}{m}{n}$: in this case, the global normalisation factor is $ 1/ \Pi_{p=1}^m\sqrt{s_p!}$ (as in \cref{eq:output}).

Following \cref{eq:efficientoutput} and \cref{eq:recstep}, we obtain the full output distribution of \sket{s}{m}{n}, along with the full output distribution of intermediary states having from $0$ to $n-1$ photons. We will denote these intermediate states as $\sket{s}{m}{k}$ and they will be stored in a list \IS of tuples: $[(\sket{0}{m}{0},p_1),\dots,(\sket{s}{m}{n-1},p_n)]$ where $\sket{s}{m}{k}$ is the Fock state $\sket{s}{m}{k-1}$ with one more photon in the mode $p_k$.

In practice, obtaining the full distribution $\Ua \sket{s}{m}{k}=\sum_{\sket{t}{}{} \in \F{m}{k}} \sbra{t}{}{} \Ua \sket{s}{m}{k} \ket{t}$ means computing and storing every coefficient 
$\sbra{t}{}{} \Ua \sket{s}{m}{k}$ for $\sket{t}{}{} \in \F{m}{k}$. We store each of them in an array, where each coefficient of $\Ua \sket{s}{m}{k}$ is used for the computation of $\Ua \sket{s}{m}{k+1}$. Note that in the sum of \cref{eq:recstep}, $\opc{a}{i} \bra{t} \Ua \sket{s}{m}{k} \ket{t} = \sqrt{t_i+1} \bra{t} \Ua \sket{s}{m}{k} \ket{t_1,\dots,t_i+1,\dots,t_m}$, which adds a new term for  $\bra{t_1,\dots,t_i+1,\dots,t_m} \Ua \sket{s}{m}{k+1}$ as in the formula of Algorithm \ref{alg:slosfull}.

\begin{algorithm}[t]

\SetKwFunction{FMain}{SLOS\_full}
\SetKwProg{Fn}{Function}{:}{}

\textbf{Global} \Ua
\\
\Fn{\FMain{$\sket{s}{m}{n},\U$}}{
$ \Ua[\ket{0}][\ket{0}] \leftarrow 1$ ;

$\IS= \left[\left(\sket{0}{m}{0},p_1\right),\dots,\left(\sket{s}{m}{n-1},p_n\right)\right]$ 
\tcp*{chosen arbitrarily, see \cref{eq:efficientoutput}}

\For{$k :0 \to n-1$}{
$\sket{s}{}{},p \leftarrow \IS{[k]}$  ;

\For{$\sket{t}{}{} \in \F{}{k} $}{
\For{$i \in [m]$}{

$\displaystyle \Ua [\sket{s_1,\dots,s_p+1,\dots,s_m}{}{}][\ket{t_1,\dots,t_i+1,\dots,t_m}] \mathrel{+}= \sqrt{\frac{t_i+1}{s_p+1}}  \times \U[i][l] \times \Ua[ \ket{s}][\ket{t}] $ ;}
\tcp{Possible memory optimization here: we don't need $\Ua[ \ket{s}][\ket{t}]$ anymore}}
}
}
\caption{
\label{alg:slosfull}
{\SLOSfull} with one input \sket{s}{m}{n} computing the full distribution \Ua\sket{s}{m}{n}.
\textit{Each coefficient $\sbra{t}{}{} \Ua \sket{s}{}{}$ will be stored in an array {\Ua} and will be accessed by $\Ua[\ket{s}][\ket{t}]$. The access of the coefficient $u_{i,p}$ will be labelled as $\U[i][p]$. Every coefficient of \Ua \ is initialised at $0$. For simplicity, the array \Ua \ is of size $\bigO{(\frac{n}{m}+1)\Mn}$, even though it could be optimised to $\bigO{\Mnl+\Mn}$, by reallocating memory and erasing the intermediary states, as shown below in the comment. 
} }
\end{algorithm}

\paragraph{Time Complexity} \SLOSfull computes the full distribution of an input \sket{s}{m}{n} in $O(n \Mn )$. The complexity can be directly deduced from the three ``for loops'' of Algorithm \ref{alg:slosfull}, knowing that $\#\F{m}{k}=\CMk=\Mk$. The total number of operations is:
$$\sum_{k=0}^{n-1} m \Mk = m \frac{n}{m}\Mn = n\Mn$$

The complexity is therefore linear in the number of states, so each state needs $O(n)$ operations in average.

\paragraph{Exponential gain from the permanent-based method}
We would now compare that complexity to the permanent-based method, which would compute every term independently using the permanent algorithm of \cite{Shche2019}. Therefore, we need to sum over all the possible outputs states the term \minperm. To simplify the expression of the sum, we will use a lower bound, considering $n\geq1$ so that $\min\limits_{t_l\neq0}(t_l+1)\leq n+1 \leq 2n$. We therefore have the following lower bound: 
$$ \frac{1}{2}\sum\limits_{\ket{t}\in \F{m}{n}} \prod_{i=1}^m(t_i+1) \leq \sum\limits_{\ket{t}\in \F{m}{n}} \minperm$$
As shown in the Lemma 2 of \cite{clifford2020faster}, the sum of the left-hand side is equal to $\binom{2m+n-1}{n}$, giving our lower bound of $\Omega\binom{2m+n-1}{n}$. To compare with the time complexity of \SLOS, we can study the ratio $\frac{\binom{2m+n-1}{n}}{nM_n}=\frac{1}{n}\frac{\binom{2m+n-1}{n}}{\binom{n+m-1}{n}}$. By assuming $m=\theta n$ and for a fixed value $\theta$, (cf Corrollary 2 of \cite{clifford2020faster}), we can apply Stirling's formula to show the ratio is $\Omega(\frac{1}{n}\rho_\theta^n)$ with $\rho_\theta=\rhotheta$, showing the exponential speedup of \SLOS. Note that $\rho_1\approx 1.69$, $\rho_2\approx1.80$, and that $\lim\limits_{\theta\to \infty}\rho_\theta=2$. 

\paragraph{Memory Complexity}  At the step $k$, we need to use a memory of at most \bigO{\Mk + \Mkp}, as we use all the coefficients of $\Ua \sket{s}{m}{k}$ to compute and store the new coefficients of $\Ua \sket{s}{m}{k+1}$. At the end of the step $k$, we can erase all the coefficients of the step $k-1$.

Therefore, at the last step, we have stored at most $\Mnl+\Mn \leq 2\Mn=\bigO{\Mn}$ coefficients.
Thus, if we allow reallocation, the complexity in memory is in \bigO{\Mn}.
As we aim for the full distribution containing $\Mn$ amplitudes, we necessarily need a memory of $\bigO{\Mn}$. 
\SLOSfull has therefore an optimal complexity in memory.

It is important to highlight for a simple allocating, we need $\Mnl+\Mn$ space which can still be inconvenient for big values of $n$ and $m$. We can reduce the overhead by only allocating the memory of $\Ua [\sket{s}{}{k+1}][\ket{t_1,\dots,t_i+1,\dots,t_m}]$ when needed, and by erasing coefficients as soon they have been used as in Algorithm \ref{alg:slosfull}. We can optimise even more by ordering \F{m}{k} so that for consecutive \sket{t}{}{}, the coefficients  $\ket{t_1,\dots,t_i+1,\dots,t_m}$ overlap and less memory is added at each iteration of a new \sket{t}{}{}. The memory optimisation can also only be done for the last steps as they are the most costly. 

For faster overhead in time and easier implementation, we would rather store all the coefficients of the intermediary states without erasing them. This would require to store $\sum_{k=0}^n\Mk=(\frac{n}{m}+1)\Mn$ states, which is still feasible for reasonable values of $m$ and $n$.

\subsection{\SLOSgen: Computation of Several Outputs for Several Inputs}

The algorithm \SLOSfull always computes the full output distribution, and does not offer any granularity by restricting the set of outputs states.
However, often we may not need the full distribution for a given input. It is the case when we are looking for one coefficient or specific coefficients, as for post-selected or heralded scheme introduced in \cref{sec2:SLOSproblems}.

\label{subsec:SLOSgen}
\subsubsection{Restriction of the Subcomputation Space: Mask }
\label{subsubsec:mask}
In order to efficiently restrict the output distribution, we introduce the notion of a \emph{mask}, a state which will filter out unnecessary intermediary states.
We define the relation $\leq$ as:
$\sket{t}{m}{} \leq \sket{\mathcal{M}}{m}{} \Leftrightarrow \bigwedge_{i=1}^m (t_i \leq \mathcal{M}_i)$, and we define $\FM{}{k} = \left\{ \sket{t}{}{} \middle|  \sket{t}{}{} \in \F{m}{k}, \sket{t}{}{} \leq \sket{\mathcal{M}}{m}{} \right\}$. For a set of masks $\mathcal{S_M}$, we will note as $\F{\mathcal{S_M}}{k}$ all the states we compute at the step $k$.

At the step $k$ of the computation, we compute each coefficient as follows:

\begin{equation} \label{eq:coefrec}
 \displaystyle 
\bra{t} \Ua \sket{s}{}{k} = \frac{1}{\sqrt{s_{p_k}}} \sum_{i,t_i \neq 0} \sqrt{t_i} u_{i,p_k}  \bra{t_1,\dots,t_i-1,\dots,t_m}  \Ua \sket{s}{}{k-1}
\end{equation}

As $\sket{t}{}{k} \in \FM{}{k} \Rightarrow \left \{\ket{t_1,\dots,t_i-1,\dots,t_m}, t_i \neq 0 \right\} \subset \FM{}{k-1} $, for any mask $\mathcal{M}$, the set $\{ \tUs^k, \ket{t} \in \FM{}{k} \}$ can be computed with $\{ \tUs^{k-1}, \ket{t} \in \FM{}{k-1} \}$.
For each $k$, it is sufficient to compute the states $\F{\mathcal{S_M}}{k} = \left\{ \FM{}{k} \middle| \mathcal{M} \in \mathcal{S_M} \right\}$.
The number of intermediary states is therefore
$\# \F{\mathcal{S_M}}{k}$
instead of \Mk for the full distribution in \cref{subsec:SLOSfull}.
Given a set of outputs $\mathcal{O}=\{ \sket{o}{m}{n}, \sket{o}{m}{n} \in \F{m}{n} \}$ we would like to compute, we can take
$\F{\mathcal{S_M}}{k}= \cup_{o\in \mathcal{O}} \F{\leq o}{k}$, so that at the last step we have $\F{\mathcal{S_M}}{n}=\mathcal{O}$.

Unlike the iterative presentation of \SLOSfull, it is here more natural to have a recursive procedure, as we can directly build the intermediary states as follows: 
\begin{equation} \label{eq:fkrec}
    \FSM{k-1}=\left\{\ket{t_1,\dots,t_i-1,\dots,t_m}, t_i \neq 0, \sket{t}{m}{k} \in \FSM{k} \right\}
\end{equation}
This makes it possible to build $\F{\mathcal{S_M}}{k-1}$ from $\F{\mathcal{S_M}}{k}$ on the fly without precompiling anything.

\subsubsection{The \SLOSgen Algorithm}
 \label{subsec:manyinputs}
 
\begin{algorithm}[tbp]
\caption{\SLOSgen with a set of $q$ inputs $\mathcal{I}$ and a set of $r$ outputs $\mathcal{O}$. \\
\textit{Each coefficient $\sbra{t}{}{} \U \sket{s}{}{}$ will be stored in a dictionary or an array \Ua and will be accessed by $\Ua[\ket{s}][\ket{t}]$. The access of the coefficient $u_{i,p}$ will be labelled as $\U[i][p]$. Every coefficient of \Ua is initialised at $0$. For simplicity, the array \Ua is of size $\bigO{q(\frac{n}{m}+1)\Mn)}$, even though it could be optimised by reallocating memory and erasing the intermediary states.}\label{alg:slosmanyinputs}}
\SetKwFunction{FRec}{SLOS\_Rec}
\SetKwFunction{FMain}{SLOS\_gen}
\SetKwProg{Fn}{Function}{:}{}

\textbf{Global} \Ua

\Fn{\FRec{$k,\FSM{k+1},\U$}}{

\If{$k>0$}{\FRec{$k-1,\FSM{k},\U$}{} \tcp*{Build with \cref{eq:fkrec}} }

\For{$(\sket{s}{}{},p) \in \IS{[k]}$}{
\For{$\sket{t}{}{} \in \FSM{k+1} $}{
\For{$i \in [m] \textbf{ when } t_i \neq 0$}{

$\displaystyle \Ua [\sket{s_1,\dots,s_p+1,\dots,s_m}{}{}][\ket{t}] \mathrel{+}= \sqrt{\frac{t_i}{s_p}}  \times \U[i][p] \times \Ua[ \ket{s}][\ket{t_1,\dots,t_i-1,\dots,t_m}] $ }}
}{

}}

\Fn{\FMain{$\mathcal{I},\mathcal{O},\U$}}{
$ \Ua[\ket{0}][\ket{0}] \leftarrow 1$ ;

$\IS{}=\left[\left\{\left(\sket{0}{m}{0},p_1^i\right)\right\},\dots,\left\{\left(\sket{s^i}{m}{n-1},p_{n}^i\right)\right\} \right]$ \tcp*{As in Algorithm \ref{alg:IS}}

\FRec{$n-1,\mathcal{O},\U$}{} ;

}
\end{algorithm}

\begin{algorithm}[tbp]
\caption{\label{alg:IS} Construction of \IS. The function  $\textbf{path}(\ket{t}\rightarrow \ket{s})$ returns an arbitrary path from $\ket{t}$ to $\ket{s}$ as in \cref{eq:efficientoutput}.}

$\IS=[\{\} \textbf{ for } i \in [m] ]$ 

\While{$\#\mathcal{I} \geq 2$}{
argmax,max $\leftarrow$ \textbf{None},0

\For{$(\sket{s}{}{},\sket{s'}{}{}) \in \left\{ (\sket{s}{}{},\sket{s'}{}{})\in \mathcal{I}^2, \ket{s} \neq \ket{s'}   \right\}$}{
$k=\sum_{i=1}^{m} \min(s_i,s'_i)$

\If{k$>$max}{
argmax,max$\leftarrow(\sket{s}{}{},\sket{s'}{}{})$ ,k

}
}

\If{max$=$0}{ \tcp{No factorisation found: we add the paths of all remaining elements in \IS and empty $\mathcal{I}$}  
\For{$\sket{s}{}{} \in \mathcal{I}$}{
\For{$(\sket{t}{m}{i},l) \in \textbf{path}(\sket{0}{m}{} \rightarrow \ket{s})$} {$\IS\text{[i]}.add((\sket{t}{m}{i},l))$}
} 
$\mathcal{I}.clear()$
}

\Else{
\tcp{Factorisation for a pair found: we add the paths of the elements of the pair in \IS, remove them from $\mathcal{I}$, and add the factor in $\mathcal{I}$}  
$(\ket{s},\ket{s'}) \leftarrow$ argmax

$\kminss \leftarrow \ket{\min(s_1,s_1'),\dots,\min(s_m,s_m')}$

\For{$\ket{s^\star} \in (\ket{s},\ket{s'})$}{
\For{$(\sket{t}{m}{i},p) \in \textbf{path}(\kminss \rightarrow \ket{s^\star})$}{$\IS\text{[i]}.add((\sket{t}{m}{i},p))$}
$\mathcal{I}.remove(\ket{s^\star})$}
$\mathcal{I}.add(\kminss)$
}
}

\If{$\#\mathcal{I} == 1$}{
\tcp{One element remaining: we add its path in \IS}

$\ket{s} \leftarrow \mathcal{I}.pop() $ 

\For{$(\sket{t}{m}{i},p) \in \textbf{path}(\ket{0}_m \rightarrow \ket{s})$}{$\IS\text{[i]}.add((\sket{t}{m}{i},l))$}
}

\end{algorithm}

Shown in Algorithm~\ref{alg:slosmanyinputs},  \SLOSgen starts with $\FSM{n}=\mathcal{O}$. The recursive computation of intermediary states is directly derived from \cref{eq:coefrec}: $\texttt{SLOS\_rec}(\FSM{k})$ computes $\{ \tUs^k, \ket{t} \in \FSM{k} \}$ from $\{ \tUs^{k-1}, \ket{t} \in \FSM{k-1} \}$ given by $\texttt{SLOS\_rec}(\FSM{k-1})$.

However, unlike the case of \SLOSfull, we are also computing the outputs of \emph{several} input states at the same time. It is in fact possible to take advantage of the similarities in the inputs, by changing the order of computation of \cref{eq:output}.

\paragraph{Several Inputs}
Let us consider two inputs \sket{s}{m}{n} and  \sket{s'}{m}{n}, having $j$ photons in common modes. Without loss of generality, we can write them as $\ket{d_1+c_1,\dots,d_m+c_m}$ and $\ket{d'_1+c_1,\dots,d'_m+c_m}$, with $d_i,d'_i,c_i \geq 0$ and $\sum_{i=1}^m c_i =j$. 
We can therefore first compute $\Ua \ket{c_1,\dots,c_m}_m=\prod_{p=1}^m (\sum_{i=1}^n u_{i,p}\opc{a}{i})^{c_p} \sket{0}{m}{}$, common term of both $\Ua\sket{s}{m}{n}$ and $\Ua\sket{s'}{m}{n}$. We can notice that $\sket{c}{m}{j}=\ket{\min(s_1,s'_1),\dots,\min(s_m,s'_m)}$ and so $j=\sum_{i=1}^m \min(s_i,s'_i)$.

More generally, with a set of inputs, we can always factorize the common terms, and use it in the computation to reduce the number of intermediary states. It is especially efficient when the common term is the same for all the inputs, as it is the case when we add ancilla photons for all different inputs, as in 
$\ket{*,*,\dots,*,1,1,1}$.

When there are many inputs, we want to build \IS{} that implies the minimum number of operations. This time, \IS{} will be a list of sets, where $\IS{[k]}$ is the set of tuples $\left( \sket{s^i}{m}{k},p_k^i \right )$ to compute at the step $k$.
Let's note $\mathcal{I}_{2}$ as the set of all distinct pairs of $\mathcal{I}^2$.
If we have $p\geq 2$ inputs, then we would choose one of the biggest common term among all pairs of $\mathcal{I}_{2}$. Therefore, we would take one element of:
\begin{equation}\label{eq:argmax}
\argmaxx_{(\sket{s}{}{},\sket{s'}{}{})\in \mathcal{I}_2} \sum_{i=1}^{m} \min(s_i,s'_i)
\end{equation}
We then remove the chosen pair $(\sket{s}{}{},\sket{s'}{}{})$ from $\mathcal{I}$, and  add $\sket{\min(s_1,t_1),\dots,\min(s_m,t_m)}{m}{}$ both to $\mathcal{I}$ and to $\IS$, with arbitrary paths from that state to $\sket{s}{}{}$ and $\sket{s'}{}{}$.
The procedure will continue until $p=1$, where we will just add the last input in $I_S$, with an arbitrary path, as in \cref{eq:efficientoutput}.
The pseudocode is the Algorithm \ref{alg:IS}.

\paragraph{Example}
To illustrate the correspondence between $\mathcal{I}$ and $\mathcal{I_S}$, we give the steps for the computing the coefficients associated to $\{ \ket{1,0,0,1},\ket{1,1,1,2},\ket{1,1,2,1} \}$. 
\begin{enumerate}
    \item This is the initial input to the procedure, and $\IS$ is empty: $\mathcal{I}=\{ \ket{1,0,0,1},\ket{1,1,1,2},\ket{1,1,2,1} \}$, 
    $ \IS=[\{\},\{\},\{\},\{\},\{\}]$
    \item
    We look for the best common factor between the three inputs (the biggest common term of \cref{eq:argmax}): it is $\ket{1,1,1,1}$, factor of $\ket{1,1,1,2}$ and $\ket{1,1,2,1}$. So now
    $\mathcal{I}=\{ \ket{1,0,0,1},$ $\ket{1,1,1,1} \}$, and $\IS$ is populated with the operations needed to recover the two elements being factored:
    $\IS=[\{\},\{\},\{\},\{\},\{(\ket{1,1,1,1},4),(\ket{1,1,1,1},3)\}]$ 
    \item
    The next step is similar, with the difference that the common factor is a term to be computed (so we keep it).
    $\mathcal{I}=\{ \ket{1,0,0,1}\}$, and
    $\IS=[\{\},\{\},\{(\ket{1,0,0,1},2)\},$ $\{(\ket{1,1,0,1},3)\},$ $\{(\ket{1,1,1,1},4),$ $(\ket{1,1,1,1},3)\}]$ 
    \item Finally, we can empty $\mathcal{I}$: $\mathcal{I}=\{\}$, and
    $\IS=[\{(\sket{0}{4}{},1)\},\{(\ket{1,0,0,0},4)\},\left\{(\ket{1,0,0,1},2)\right\},$ $\{(\ket{1,1,0,1},3)\},$ $\{(\ket{1,1,1,1},4),$ $(\ket{1,1,1,1},3)\}]$
\end{enumerate}


\begin{table}
    \centering
        \begin{tabular}{|c|c|c|c|c|}
        \hline
         & \multicolumn{2}{c|}{Time Complexity}  & \multicolumn{2}{c|}{Memory Complexity}\\\hline  $\#$Outputs & One & All & One & All\\ \hline
         Permanent-Based \cite{Shche2019} & 
           $\bigO{\minperm}$ &$\Omega\binom{2m+n-1}{n}$&$\bigO{n}$&$\bigO{\Mn}$  \\\hline 
         \SLOSgen & \boldmath${\bigO{\sumprodperm}}$ &\boldmath${\bigO{n\Mn}}$&$ O\binom{n}{n/2}$&\boldmath${\bigO{\Mn}}$ \\\hline
    \end{tabular}
\caption{Time and memory complexity analysis of \SLOSgen and \cite{Shche2019} for a generic output $\ket{t}=\ket{t_1,\dots,t_m}\in\F{m}{n}$, where $\Mn=\#\F{m}{n}=\binom{m+n-1}{n}$. We highlighted the \SLOS values in bold when they were similar or better than the permanent-based ones. For the memory complexity of the one output case, we consider the worst case, as detailed in \ref{subsubsec:mem}. For the memory complexity of the full distribution case, we consider the problem of storing every value, therefore needing at least $\bigO{\Mn}$ of memory. If we consider the enumerating problem instead, the permanent-based method would have a memory complexity of  $\bigO{n}$, far better than the $\bigO{M_n}$ of \SLOS. It is an open problem to how much the memory efficiency can be increased.}
\label{tab:complexity}
\end{table}

\paragraph{Complexity of \SLOSgen}
The theoretical complexity of the algorithm \SLOSgen is difficult to assess because it heavily depends on the redundancy in input and output states. In \cref{sec:limit-case-one-one} we analyse the limit case where only one input and one output are considered, and how it relates to the \naive method of \cite{Shche2019}. The results are summarised in the Table \ref{tab:complexity}. In \cref{sec:implem}, we present a concrete implementation of \SLOSgen in the general case, and we discuss concrete benchmarks.

\subsubsection{Limit Case with One Output/One Input}
\label{sec:limit-case-one-one}

The algorithm \SLOSgen can be specialized to the case where one only consider one input and one output state. 
This is the typical case that can be directly handled by \cite{Shche2019}: it consists in computing \cref{eq:permcoef}, so one can rely on the complexity results for computing the permanent of a matrix with potential repeated row or columns.

In this section, we discuss the complexity of the procedure in this simple one-input, one-output case, and compare it with \cite{Shche2019}. 
\paragraph{Conjugate trick}
While it is usually more physical to have one photon at most per mode in the input $\sket{s}{m}{n}$, if \sket{s}{m}{n} has more repetition than the output \sket{o}{m}{n}, it is faster and equivalent to compute $\sbra{s}{}{} \Ua^\dag \sket{o}{}{\dag}$ than $\sbra{o}{}{} \Ua \sket{s}{}{}$. That trick of taking the conjugate transpose allows to transform the repetitions of the columns into the repetition of the rows if necessary, so the number of computation is reduced as much as possible. However, that trick is not general, as it is for a specific instance of one output, and needs more repetitions in the input, that is not very likely in practice with single photon sources. In the following, we will consider a generic input $\ket{s}$ and only consider the repetitions in the output $\ket{t}$.
\label{subsec:oneoutonein}
\paragraph{Worst Case}
To efficiently compute the output $\bra{o} \Ua \ket{s}^n_m$ , one has to take the mask  $\sket{\mathcal{M}}{m}{}=\sket{o}{m}{n}$.
The worst case is when the mask is of the form $\ket{1,1,\dots,1,0,\dots,0}^n$, as it maximises the size of \FM{}{k} for each $k$. In that case, $\# \FM{}{k} = \binom{n}{k}$, and each term needs $k$ operations, as it is is the number of terms in the sum of \cref{eq:coefrec}.
Therefore, the number of needed operations is $\sum_{k=1}^n k \binom{n}{k} = n2^{n-1}$.
The complexity therefore matches the time complexity for computing a general permanent, as $\bra{1,\dots,1,0,\dots,0} \Ua \ket{1,\dots,1,0,\dots,0} $ is computed by the permanent of a matrix without any repetition of rows or columns.

\paragraph{General Case} For each state $\sket{t}{m}{} \in \F{m}{}$, let's define $\alphat=\#\left\{t_i\neq0,  1\leq i\leq m\right\}$. From Eq. \ref{eq:coefrec}, we can see that every $\bra{t} \Ua \sket{s}{}{k}$ needs $\alphat$ computations, using the already computed coefficients. Given an output state $\ket{t}$, we will compute every $\ket{t'}\leq \ket{t}$ 
 where $\leq$ is defined in \ref{subsubsec:mask}. Therefore, the number of computations is $S(\ket{t})=\sum\limits_{\ket{t'}\leq\ket{t}}\alphatp$. For the proof, we will use the notation $\ket{t}_k=\ket{t_1,\dots,t_k}$, restriction of $\ket{t}_m=\ket{t}$. To give a generic formula of $S$, we proceed by induction by defining $S_k(\ket{t})=\sum\limits_{\ket{t'}\leq\ket{t}_k}\alphatp$, with $S_m(\ket{t})=S(\ket{t})$ and $S_1(\ket{t})=t_1$. We can show that
\[\begin{array}{rcl}  S_m(\ket{t})&=& \sum\limits_{\ket{t'}\leq\ket{t},t'_m=0}\alphatp + \sum\limits_{k=1}^{t_m}\sum\limits_{\ket{t'}\leq\ket{t},t'_m=k}\alphatp \\ 
&=& S_{m-1}(\ket{t}) + \sum\limits_{k=1}^{t_m}\sum\limits_{\ket{t'}\leq\ket{t}_{m-1}}(\alphatp+1)\\
&=& S_{m-1}(\ket{t}) + t_mS_{m-1}(\ket{t}) + t_m\sum\limits_{\ket{t'}\leq\ket{t}_{m-1}}1\\
&=& (t_m+1)S_{m-1}(\ket{t}) + t_m \prod\limits_{i=1}^{m-1}(t_i+1) 
\end{array}\]
Leading to the following formula: 
\[\begin{array}{rcl}S(\ket{t})&=&\sumprodpermlim\end{array}\]
We can show that it is always less than the bound found in \cite{Shche2019}:
\[\begin{array}{rcl}\sumprodpermlim &=& \sum\limits_{t_i\neq0} t_i\prod\limits_{j\neq i}(t_j+1)\\ &\leq& \sum\limits_{t_i\neq0}t_i \frac{1}{\min\limits_{t_l\neq0}(t_l+1)}\prod\limits_{j=1}^m(t_j+1) \\ &=& \left(\frac{\prod_{j=1}^m(t_j+1)}{\min\limits_{t_l\neq0}(t_l+1)} \right)\sum\limits_{t_i\neq0}t_i \\ &=& n \left(\frac{\prod_{j=1}^m(t_j+1)}{\min\limits_{t_l\neq0}(t_l+1)} \right) \end{array}\]
That bound is strict whenever there are two $t_i>0,t_j>0$ such that $t_i\neq t_j$.

\paragraph{Average Case}
For each state $\sket{t}{m}{k} \in \F{m}{k}$, for each $u_{i,*}$ with $i \in [m]$, the term   $u_{i,*}\bra{t}\Ua\ket{s}^k$ will be used as many times as $\ket{t_1,\dots,t_i+1,\dots,t_m}^{k+1} \leq \sket{\mathcal{M}}{m}{n}$ with $\sket{\mathcal{M}}{m}{n} \in \F{m}{n}$. We can show that the number of time is $M_{n-k-1}$, which is the number of putting $n-k-1$ photons into $m$ modes, as $\ket{\mathcal{M}}$ is necessarily of the form $\ket{t_1+q_1,\dots,t_i+1+q_i,\dots,t_m+q_m}$ with $q_i\geq 0$ and $\sum_i q_i =n-k-1$.

As there are $\Mn$ different output states, we can show the average number of operation for computing one output is:

$$ \frac{1}{ \Mn} \sum_{k=0}^{n-1} m \Mk M_{n-k-1} =\frac{m}{\Mn} \displaystyle \binom{2m+n-2}{n-1}=\frac{n}{M_n}\binom{2m+n-1}{n}\frac{m}{2m+n-1}$$

It is therefore the same average complexity of \cite{clifford2020faster} (cf. the Corollary 3 for the precise formula), with a multiplicative ratio of $\frac{m}{2m+n-1}<1$. 

\paragraph{Memory Complexity} 
\label{subsubsec:mem}
Even if \SLOSgen is better in time for one input and one output than \cite{Shche2019}, it is important to highlight it is far more costly in memory. Given an output state $\ket{t}^n$, at the step $k$, we need to store all the coefficients $\bra{t}\Ua\ket{s}^{k-1}$ and $\bra{t}\Ua\ket{s}^{k}$, with $\ket{t}^{k-1}\leq \ket{t}^k\leq \ket{t}^n$. As $\# \left\{\ket{t}^k \leq \ket{t}^n \right\}$ doesn't have a simple formula \cite{stack}, we will only consider the worst case. The worst case is when $\ket{t}$ has at most one photon per mode, for instance when $\ket{t}=\ket{1,1,\dots,1,0,\dots,0}^n$. In that case, the number of space needed at each step is $\bigO{\binom{n}{k}}$. As we only need to store the values of two steps, and not all the $n$ steps, we can only consider the costliest which is when $k=n/2$. The worst case space complexity is therefore $\bigO{\binom{n}{n/2}}$.

\begin{algorithm}[t]  \caption{\label{alg:sloshyb}\SLOShyb with one input \sket{s}{m}{n} and \Select function. \\
\textit{Each coefficient $\sbra{t}{}{} \U \sket{s}{}{}$ will be stored in a dictionary or an array \Ua and will be accessed by $\Ua[\ket{s}][\ket{t}]$. The access of the coefficient $u_{i,p}$ will be labelled as $\U[i][p]$. Every coefficient of \Ua \ is initialised at $0$. For simplicity, the array \Ua \ is of size $\bigO{(\frac{n}{m}+1)\Mn}$. The \Select function enables to do Strong simulation without or with masks, Weak simulation, or a mix of them. \Ba{} is an array or dictionary initialised at False and stores the coefficients already computed, so we don't need to recompute $\Ua[\ket{s}][\ket{t}]$ if $\Ba[\ket{s}][\ket{t}]$ is true. $\FM{}{k} \cap \mathcal{\overline{B}}_s$ is a shortcut for $\{ \ket{t}| (\ket{t} \in \FM{}{k}) \wedge (\Ba[\ket{s}][\ket{t}]==False)  \}$  } }

\SetKwFunction{FMain}{SLOS\_hyb}
\SetKwProg{Fn}{Function}{:}{}
\SetKwFunction{FRec}{SLOS\_Rec}
\SetKwFunction{FSel}{Select}

\textbf{Global} \Ua, \Ba{}

\Fn{\FRec{$k,\IS,\F{S}{k},\U$}}{

$(\sket{s}{}{},p) \leftarrow \IS{[k]}$

\If{$k>1$}{\FRec{$k-1,\IS,\F{S}{k-1} \cap \mathcal{\overline{B}}_s,\U$}{} \tcp*{Build with \cref{eq:fkrec}} }

\For{$\sket{t}{}{} \in \F{S}{k} $}{
\For{$i \in [m] \textbf{ when } t_i \neq 0$}{

$\displaystyle \Ua [\sket{s_1,\dots,s_p+1,\dots,s_m}{}{}][\ket{t}] \mathrel{+}= \sqrt{\frac{t_i}{s_p}}  \times \U[i][p] \times \Ua [ \ket{s}][\ket{t_1,\dots,t_i-1,\dots,t_m}] $ }$\Ba[\ket{s}][\ket{t}] \leftarrow$ True}
}{}
\Fn{\FMain{$\sket{s}{m}{n},\U,\Select$}}{

$ \Ua[\ket{0}][\ket{0}] \leftarrow 1$ ;

$\IS= \left[\left(\sket{0}{m}{0},p_1\right),\dots,\left(\sket{s}{m}{n-1},p_{n}\right)\right]$ 
\tcp*{chosen randomly from \cref{eq:output}}
$\F{S}{k}=\{ \sket{0}{}{} \}$

\For{$k:1 \to n$}{
$\F{S}{k} \leftarrow \Select(\IS,\F{S}{k},k)$ 

\FRec{$k,\IS,\F{S}{k},\U$}

}
return $\F{S}{k}$ }
\end{algorithm}

\subsection{\SLOShyb: General Procedure for both Weak and Strong Simulations}
\label{subsec:SLOShyb}

This section presents \SLOShyb (shown in Algorithm~\ref{alg:sloshyb}), a hybrid of weak and strong simulation. It can be seen as an extension of \SLOSgen, with an iterative procedure and a recursive subroutine.

Let us denote $\F{S}{k}$ the set of intermediary states computed at the step $k$ with \cref{eq:coefrec}. In \cref{subsec:SLOSfull} we were interested in computing every states, so $\F{S}{k}=\F{m}{k}$, while in \cref{subsec:SLOSgen} we had masks to compute specific outputs, so $\F{S}{k}=\FSM{k}$. For the hybrid version in Algorithm \ref{alg:sloshyb}, we use a \Select function to select the next intermediary states we would like to compute, so $\F{S}{k+1}$ is defined as $\Select (\F{S}{k})$. In order to compute all coefficients in $\F{S}{k+1}$, we use \texttt{SLOS\_Rec} with the particularity to only compute intermediary states that have not been computed so far, and we directly use the intermediary states needed for the coefficients of \F{S}{k}. We can code that information in a Boolean array (or a hash table) \Ba{} such that \Ba{[\ket{s}][\ket{t}]} is true if $\Ua[\ket{s}][\ket{t}]$ has been computed, false otherwise.

For strong simulations, we are aiming at computing the probabilities of a given set of outputs. However, sometimes we would like to compute the states with the biggest probabilities, without knowing in advance the probability distribution of the set of outputs. More generally, we would like to compute outputs giving conditions on the distribution even if we don't know it yet.
For instance, \SLOShyb can solve the following problem.

\paragraph{Problem 3: Hybrid Simulation}

Given an input \sket{s}{m}{n}, a threshold $\eta$ and a LO-circuit of $m$ modes mapping a unitary \U$:m\times m$, can we obtain a set $\mathcal{O}=\left \{ \sket{t}{}{}, \sket{t}{}{} \in \F{m}{n} \right \}$ such that $\sum_{\ket{t} \in \mathcal{O}} \amp{\bra{t}\Ua\ket{s}}^2 > \eta $?

Even though this problem can be solved with \ref{prob1} and just taking a subset of the full distribution, we can have a more efficient procedure by using the Algorithm~\ref{alg:sloshyb} and the \Select function of Algorithm~\ref{alg:select}.
Note the ``while loop'' adds several states in \F{}{}.

\paragraph{Weak Simulation}The algorithm of \cite{clifford_classical_2018} uses the probability chain rule to sequentially sample photon by photon. At each step, some probabilities are computed in order to sample the outputs $\ket{t}^k$ for $1\leq k \leq n$. At the end, we sample the desired sample $\ket{t}^n$ with $n$ photons from the desired distribution. To lighten the notation, we will note $t^k=\ket{t}^k,s^k=\ket{s}^k$, and we will note $p(t^k|t^{k-1},s^k)$ the conditional probability to sample $\ket{t}^k$ conditioned on $\ket{t}^{k-1}\leq \ket{t}^k$, meaning we know the position of $k-1$ photons in the output, and conditioned on the input\footnote{In the case of Boson Sampling we generally have $\ket{s}^n = \ket{1,1,1,\dots,1,0,\dots,0}_m$.} being $\ket{s}^k$. \\
The main idea is to avoid directly sampling from $p(t^k)$, by noticing we can equivalently sample from $p(t^k|s^k)$, where the inputs $\ket{s}^k$ are randomly chosen. Then, to sample from $p(t^k|s^k)$, we use the chain rule to sample from $p(t^k|t^{k-1},s^k)$ instead, given we already sampled $\ket{t}^{k-1}$. The technical points can be summarised as follows\footnote{We use $\propto$ to lighten the formula and disposing of factorials coefficients depending on $k,n$.}:
\begin{itemize}
\item We can show that $p(t^k)\propto \sum\limits_{\ket{s}^k\leq\ket{s}^n} p(t^k|s^k)\propto\mathbb{E}_{\ket{s}^{k}\leq\ket{s}^{n}}\left\{p(t^{k}|s^{k})\right\}$, with $p(t^k|s^k)=\amp{\bra{t}\Ua\ket{s}^k}^2$, and with the expectation value summing uniformly over  $\ket{s}^{k}\leq\ket{s}^{n}$. (cf Lemma 1 and 3 of \cite{clifford_classical_2018}\footnote{In that paper, the \text{photon to mode} encoding as described in Section \ref{sec:memory} is used, so the positions of the $i^{th}$ photon is described by $r_i\in[m]$. Linking the notations, we have $p(r^k|r_1,\dots,r_{k-1})=p(t^k|t^{k-1})$. Their {\boldmath${\alpha}$} is a way to parameterise the permutation of the columns, representing the order of the input photons. Therefore considering a random permutation $\alpha_1,\dots,\alpha_n$ is equivalent to consider a random order $\left\{\ket{s}^k, 1\leq k \leq n \right\}$ of the inputs.}). 
\item We sample from $\mathbb{E}_{\ket{s}^{k}\leq\ket{s}^{n}}\left\{p(t^{k}|s^{k})\right\}$ instead than sampling from $p(t^k)$. As the $\ket{s}^k$ are always uniformly distributed, we can first uniformly sample the order $\left\{\ket{s}^k, 1\leq k \leq n \right\}$, and then sample from the $p(t^{k}|s^{k})$ distribution. 
\item To sample from $p(t^{k}|s^{k})$, we use the chain rule. For a fixed order $\left\{\ket{s}^k, 1\leq k \leq n \right\}$ - randomly sampled at the step 2 - we have  $p(t^n|s^n)=p(t^1|s^1)p(t^2|t^1,s^2)\dots p(t^n|t^{n-1},s^{n})$. At the step $k$, given the previous sample $\ket{t}^{k-1}$, we sample the output $\ket{t}^k$ from $p(t^k|t^{k-1},s^k)$. 
\item $\left\{p(t^k|\ket{t_1,\dots,t_m}^{k-1},s^k), \ket{t}^k \geq \ket{t}^{k-1}\right\}=\left\{ \amp{\bra{t_1,\dots,t_i+1,\dots,t_m}\Ua\ket{s}^k}^2, 1\leq i\leq m \right\}$. We therefore compute those $m$ permanents to compute the desired probabilities and sample $\ket{t}^k$.
\end{itemize}
We will now explicit how \SLOShyb in Algorithm~\ref{alg:sloshyb} can perform the weak simulation. To be clearer and slightly more efficient, the commands in the for loop are inversed: we first compute the amplitudes with \texttt{SLOS\_Rec}, and then we would use the \Select function to sample: 
    \begin{enumerate}
        \item Given an input $\ket{s}^n$, 
        the random order $s^1\leq s^2 \leq \dots \leq s^n$ is done by the random choice of \IS. We initialise $\F{S}{k}=\left\{\ket{1,0,\dots,0}_m,\dots,\ket{0,\dots,0,1}_m\right\}$.  
    \item The \Select function at the step $1\leq k \leq n$ and given a set of states $\F{S}{k}$, will sample $\ket{\alpha}^{k}$ from $\left\{ \amp{\bra{t} \Ua \ket{s}}^2 \middle| \ket{t} \in \F{S}{k} \right\}$. Then if $k<n$, it will return the next values to compute, i.e. the set $\left\{\ket{\alpha_1,\dots,\alpha_i+1,\dots,\alpha_m}, i \in [m] \right\}$. If $k=n$, then we are at the last step, so we just can return the last sample $\{\ket{\alpha}^n\}$.
\end{enumerate}

\begin{algorithm}[t]

\SetKwFunction{FMain}{SLOS\_gen}
\SetKwProg{Fn}{Function}{:}{}
\SetKwFunction{FRec}{SLOS\_Rec}
\SetKwFunction{FSel}{Select}

\Fn{\FSel{$\IS,\F{S}{k},k$}}{
S $\leftarrow 0$

$(\sket{s}{}{},\star) \leftarrow \IS{[k]}$

$\F{}{} \leftarrow \{  \} $

\While{S $\leq \eta$}{ 
$\ket{t} = \F{S}{k}.pop()$

S$\mathrel{+}= \amp{\bra{t} \Ua \ket{s}}^2$

$\F{}{}.add(\ket{t})$
}
return $\left\{\ket{t_1,\dots,t_i+1,\dots,t_m}, \ket{t} \in \F{}{}, i \in [m] \right\}$ 
}
\caption{\Select function which answers to the Problem 3 \label{alg:select}}
\end{algorithm}


\section{Implementation}
\label{sec:implem}

In this section, we discuss the concrete implementation of \SLOSgen, as found in Perceval~\cite{heurtel_perceval_2022}.

To build-up intuition, \cref{tab:order} gives some orders of magnitude to the quantities involved and \cref{tab:equivalency} gives a quick equivalence of these quantities in time and storage. Considering the exponential number of values we are dealing with, our challenge is to optimize as much as possible the memory structures and the details of the implementation to gain practical factors allowing to extend the strong simulation for a few more photons.

\begin{table}
    \centering
        \begin{tabular}{|c|c|c|c|}
        \hline
        $n$ & $M_n$ for $m=n$ & $M_n$ for $m=2n$ & $M_n$ for $m=3n$ \\
        \hline
        $12$ & $1.35\times10^6$ & $8.34\times10^8$ &	$5.23\times10^{10}$ \\
        \hline
        $14$ & $2,01\times 10^7$ & $3.52\times 10^{10}$ &  $4.35\times 10^{12}$ \\
        \hline
        $16$ & $3.01\times10^8$ & $1.5\times10^{12}$ &	$3.66\times10^{14}$ \\
        \hline
        17 & $1.17\times 10^9$ & $9.85\times 10^{12}$ & $3.37\times 10^{15}$\\
        \hline
        $18$ & $4.54\times 10^9$ & $6.46\times 10^{13}$ & $3.11\times 10^{16}$\\
        \hline
        $20$ & $6.89\times10^{10}$ & $2.79\times10^{15}$ & $2.65\times10^{18}$\\
        \hline
        \end{tabular}
    \caption{Value of $M_n$ for practical combinations of $m$ and $n$. 
    }
    \label{tab:order}
\end{table}

\begin{table}[t]
    \centering
        \begin{tabular}{|c|c|c|c|}
        \hline
        & \shortstack[l]{time necessary for single\\instructions processed on\\a 1GHz computer}&\shortstack[l]{equivalent storage\\size\\(any unit)}&\shortstack{number of bytes\\necessary to store\\a pointer}\\
        \hline
        $10^6$ & 1 milliseconds & 0.95 Mega & 3\\
        \hline
        $ 10^8$ & 0.1 seconds & 95 Mega & 4\\
        \hline
        $10^{10}$ & 10 seconds & 9.3 Giga & 5\\
        \hline
        $10^{12}$ & 17 minutes & 903 Giga & 5\\
        \hline
        $10^{14}$ & \color{orange}27 hours & \color{orange}90 Tera & 6\\
        \hline
        $10^{16}$ & \color{red} 115d & \color{red} 8.8 Peta & 7\\
        \hline
        $10^{18}$ & \color{red}31 years & \color{red} 888 Peta & 8\\
        \hline
        \end{tabular}
    \caption{Equivalency of $10^k$ with concrete time and storage references}
    \label{tab:equivalency}
\end{table}

In this section we assume that we have a fixed number of modes $m$. We will first describe the memory structure we have been introducing for the implementation of the algorithms.  We then discuss the implementation challenges and the optimization tricks, and finally add a note on masking.

\subsection{Memory Structures}

The two main memory structures will be labeled as {\tt fsarray} (Fock State Array) and {\tt fsimap} (Fock State Inverse Map). {\tt fsarray} is mainly used to get indices used for building {\tt fsimap}.
Both structures depend only on $m$ and $n$ and are independent of the unitary matrix. During the actual simulation they will be used as read-only structures containing the ``{\it calculation path}". It is thus possible to pre-compute these structures and serialize them to files. For the largest ones, it is also possible to use the files as a memory-map.

\paragraph{{\tt fsarray} - Indexes of the Fock states}
Each $\mathtt{fsarray}(k)$ is an array in charge of representing all the Fock states \sket{s}{m}{k} and assigning a unique ID to each of them. We want this structure to be searchable, and to have the least memory footprint. For this we use sequences $S(\sket{s}{}{k})=(p_1,...,p_k)$ where $p_i$ is the position of the $i^{\text{th}}$ photon in the Fock state $\sket{s}{}{k}$. 
For instance: $S(\ket{1,0,0,2})=(1,4,4)$, which we denote {\tt ADD}, mapping each number to an uppercase letter for the sake of readability.

This representation of Fock states, which we call \emph{photon to mode} encoding, has the following nice properties:
\begin{enumerate}
    \item It is ordered using an intuitive lexicographical order on the $(p_1,...,p_k)$: e.g. 
    \[
    S(\ket{1,0,0,2})= {\tt ADD}  > S(\ket{1,1,1,0})={\tt ABC}.
    \]
    Thus it is searchable.
    \item The sequence $S(\sket{s}{}{})$ is iterable: $(p_1,...,p_k) \xrightarrow{+1} (p'_1,...,p'_k)$, where we can simply increment as follows: (a) $p'_k=p_k+1$ if $p_k < m$, otherwise (b) we calculate $p_{k-1}$ the same way (possibly going to $p_{k-2}, \dots, p_1$) and $p_k=p_{k-1}$, e.g.\ $(1,4,4)/\mathtt{ADD}\xrightarrow{+1}(2,2,2)/\mathtt{BBB}$. 
    \item The sequence $S(\ket{s})$ can be represented in memory with a fixed $\lceil \log_2(m) \rceil$-bit buffer.
    \item It is more compact than {\it mode to photon} encoding for $n<m$. This case is the most common use case with circuits with single photon sources, as each input mode would have at most one photon.
\end{enumerate}
Properties 1 and 3 imply that, by binary search, it is possible to find a specific Fock state in a {\tt fsarray} in $\bigO{\log_2 M_k}$.
Property 2 implies that it is even possible to not store the {\tt fsarray} if we are interested in {\bf all} the Fock states since we can directly iterate through them. Thus, we will only store $\mathtt{fsarray}(n)$, as the computation of the states $\mathtt{fsarray}(k+1)$ will only need the iteration of $\mathtt{fsarray}(k)$ and the structure {\tt fsimap} as detailed below.

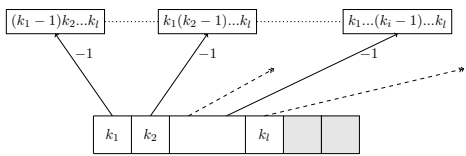
\begin{figure}
    \centering
    \scalebox{0.5}{
    \begin{tikzpicture}
    	\node (0) at (-8, 2) {};
    	\node (1) at (-6, 2) {};
    	\node (2) at (-4, 2) {};
    	\node (4) at (-8, 1) {};
    	\node (5) at (-6, 1) {};
    	\node (6) at (-4, 1) {};
    	\node (k1) at (-7.5, 1.5) {$k_1$};
    	\node (k2) at (-6.5, 1.5) {$k_2$};
    	\node (kl) at (-3.5, 1.5) {$k_l$};
    	\node[draw] (p1) at (-5, 4.5) {$k_1(k_2-1)...k_l$};
    	\node[draw] (p0) at (-9, 4.5) {$(k_1-1)k_2...k_l$};
    	\node[draw] (p2) at (0, 4.5) {$k_1...(k_i-1)...k_l$};
    	\node (28) at (-7, 2) {};
    	\node (29) at (-7, 1) {};
    	\node (30) at (-3, 2) {};
    	\node (31) at (-3, 1) {};
    	\node (7) at (-2, 1) {};
    	\node (3) at (-2, 2) {};
    	\node (45) at (-1, 1) {};
    	\node (46) at (-1, 2) {};
    	\node (34) at (-7.5, 2) {};
    	\node (35) at (-6.5, 2) {};
    	\node (36) at (-3.5, 2) {};
    	\node (37) at (-2.5, 2) {};
    	\node (38) at (-4.5, 2) {};
    	\node (39) at (-5.5, 2) {};
    	\node (43) at (1.75, 3.25) {};
    	\node (44) at (-3.25, 3.25) {};
    	\draw (1.center) to (2.center);
    	\draw (2.center) to (6.center);
    	\draw (6.center) to (5.center);
    	\draw (5.center) to (1.center);
    	\draw (1.center) to (0.center);
    	\draw (0.center) to (4.center);
    	\draw (4.center) to (5.center);
    	\draw (2.center) to (30.center);
    	\draw (6.center) to (31.center);
        \filldraw[fill=gray!20!white, draw=black]
            (30.center) -- (3.center) -- (7.center) -- (31.center) -- cycle;
        \filldraw[fill=gray!20!white, draw=black]
            (45.center) -- (7.center) -- (3.center) -- (46.center) -- cycle;
    	\draw (28.center) to (29.center);
    	\draw [thick, dotted] (p0.east) to (p1.west);
    	\draw [thick, dotted] (p1.east) to (p2.west);
    	\draw [->] (34.center) -> (p0.south) node[near end,right] {$-1$};
    	\draw [->] (35.center) -> (p1.south) node[near end,right] {$-1$};
    	\draw [->] (38.center) -> (p2.south) node[near end,right] {$-1$};
    	\draw [->>, dashed] (36.center) to (43.center);
    	\draw [->>, dashed] (39.center) to (44.center);
    \end{tikzpicture}}
    \caption{\label{fig:my_label}Representation of a cell of a $\mathtt{fsimap}(n)$: here the $n$ photons are occupying $l$ mode, so we have $n-l$ pointers not used}
    
\end{figure}

\paragraph{{\tt fsimap} - Mapping between Fock States}
Our algorithms require that we store a map between the ``child'' Fock states of $k+1$ photons \sket{s}{m}{k+1} to the ``parent'' Fock states of $k$ photons \sket{s}{m}{k} (Figure \ref{fig:my_label}). We call this structure $\mathtt{fsimap}(k+1)$ and we need to build a list of $n$ {\tt fsimap}, $\mathtt{fsimap}(n)$, $\mathtt{fsimap}(n-1)$, ...,  $\mathtt{fsimap}(1)$.

For a Fock state to find its parent, we just need to remove one photon. For instance for {\tt ADD} we just remove a {\tt A} or a {\tt D} to reach {\tt AD} or {\tt DD}. 
So more generally for each $M_{k+1}$ Fock state in $\F{m}{k+1}$ we need to store up to $k+1$ indexes to the parent. Most of the time we could store less, but we prefer keeping direct access to each entry in the {\tt fsmap}, so we are keeping for each state exactly $k+1$ pointers.

The size of a layer is then: $S_{\mathtt{fsimap}(k)}=\mathtt{fsimap}(k)$ is $\lceil \log_{256}(M_{k-1}) \rceil\times k$ bytes.

So to summarize - starting from all the states in $\F{m}{k+1}$, we can find their index $idx$ thanks to $\mathtt{fsa}(k+1)$. With the index, we can find the location of the index to the parent states in $\mathtt{fsa}(k)$ by direct memory access in $\mathtt{fsimap}(k+1)$ to the position: $idx\times (k+1)\times S_{\mathtt{fsimap}(k)}$\footnote{Note that we can get a more compact version of the $\mathtt{fsimap}(k)$ for $k< n$ when the size of the index to each state is 4-bytes long. Indeed in that case, we can then get rid of the non-used cells for the parents pointers. We then lose the ability the direct access through the $\mathtt{fsa}(k)$ index, but this index is only used when referred to by child structure, so we can instead replace the index by the direct pointer to the position of the state in memory.}.

The construction of {\tt fsimap} is straightforward using the 2 involved {\tt fsarray}.

\subsection{Implementation Optimization}

All the implementation is done in C++ optimized with SIMD vectorization. Some specific points:

\begin{enumerate}
    \item \underline{Vectorization}: Cross-platform SIMD vectorization is done through MIPP \cite{cassagne2018mipp} library\footnote{https://github.com/aff3ct/MIPP} supporting SSE, AVX, AVX-512 and ARM NEON (32-bit and 64-bit) instructions. The main operation that we are interested in is ``horizontal vector sum of complex number multiplication''. This operation is not a primitive and is decomposed into the 3 MIPP primitives: {\tt interleave}, {\tt cmul}, {\tt hadd}.
    \item The \SLOSfull algorithm is fully iterative: for $k=1$ to $n$ the coefficients corresponding to $\sket{s}{m}{k+1}$ are computed in one single loop by iterating on the $\mathtt{fsimap}(n)$ structure. Each new coefficient is a sum of products of unitary matrix coefficient times coefficient of $\sket{s}{m}{k}$. There is no overhead in the implementation.
    \item \underline{Recursion}: The \SLOSgen algorithm is recursive. Starting from the index retrieved by binary search on {\tt fsaarray} on layer $n$, the expected coefficient (the probability amplitude) is calculated by recursively retrieving the coefficients for all the parents at level $k-1$. To reduce overhead caused by recursion, a bit vector is used to check if the parent coefficient is not yet calculated before going recursive. Also, all the structures necessary for Vectorization are pre-allocated globally to reduce overhead created by dynamic allocation on the stack.
    \item \underline{Memory access}: one major challenge for \SLOS algorithms resides in memory allocation. When pushed to the maximum (say, 17 photons for 34 modes) - memory structures will use several hundreds/thousands of gigabytes in memory. Each coefficient calculation needs parent coefficient scattered on this memory range limiting possibility of processor to benefit from hardware $L_x$ memory caching. Compared to local permanent calculation algorithm, this memory access time creates overhead depending strongly on hardware configuration as seen in \cref{sec:performance}.
    \item \underline{Multithreading}:
    \begin{itemize}
        \item Algorithm \SLOSfull can be fully run on multiple threads without additional overhead: for each $k$, we simply divide the coefficients list by the number of threads.
        \item For algorithm \SLOSgen, multi-threading is more limited: it is only possible to distribute top branches of the calculation.
    \end{itemize}
    In both cases, impact of the multi-threading is actually limited by \underline{memory access}: adding more threads on very large memory structure do not significantly increase the performance.
\end{enumerate}

\subsection{Masking}

Let us note that the implementation proposed is fully compliant with the notion of ``masking'' introduced in section \ref{subsubsec:mask}: technically masking is a transversal optimization allowing to reduce (potentially massively) the complete Fock state space and therefore reducing proportionally the time and storage.  

The only location in the code impacted by {\tt masking} is on the {\tt fsaarray} construction. To find all the masked items, we are keeping the same global iterations on all possible Fock states, but are skipping the ones that are not compliant with the mask. It is likely that we could find a faster iteration method on masked Fock states, however, since this only impacts pre-computing, we don't need a special optimization.

Once the pre-computed memory structures detailed below are built, there is no difference in implementation when simulating on a masked or non-masked system.  


\section{Performance}
\label{sec:performance}
In this section, we discuss practical aspects of SLOS. We
first show a typical use case demonstrating the
need of strong simulations and effectiveness of our implementation in
\cref{subsec:QML}. We then discuss possible limitations. In \cref{sec:performance_single}, we compare
performance of \SLOS for a single output with at most one
  photon per mode compared to Glynn's algorithm, state of the
art algorithm for computing the permanents of generic
matrices. Finally, in \cref{sec:memory} we analyze the space limitation of strong simulation using our algorithm.

\subsection{A Typical QML Application Requiring Strong Simulation}
\label{subsec:QML}

Using the framework Perceval \cite{heurtel_perceval_2022}, we have implemented\footnote{Details of the algorithm can be found at \url{https://perceval.quandela.net/docs/notebooks/Differential\%20equation\%20solving.html}.} the simulation of \cite{gan2021fock}, using a generic $m\times m$ interferometer and its full output probability distribution to train a model solving differential equations. For an evolving configuration of the circuit, the algorithm has to iterate on all output states \sket{t}{m}{n} corresponding to input state $\ket{1,1,...,1}_n$. Based on Broyden–Fletcher–Goldfarb–Shanno (BFGS) optimiser \cite{fletcher2013practical}, the algorithm converges in 200-400 iterations, each of them needing thousands of full distribution computations. \cref{tab:perf_qml} shows the evolution of $M_n$ for different values of $n$ and the time necessary for 150 iterations of the algorithm comparatively with the direct calculation of $M_n$ permanent with Glynn's algorithm and \SLOSfull algorithm. Use of \SLOSfull practically allows pushing of simulation from 6 photons (processed in 2min with \SLOSfull and 11h without) to 10 photons.

\begin{table}[t]
    \centering
\begin{tabular}{|l|c|c|c|c|c|c|c|c|c|c|}
\hline
 & \multicolumn{9}{c}{Number of photons $n$}& \\
\cline{2-11}
& 2 & 3 & 4 & 5 & 6 & 7 & 8 & 9 & 10 & 11\\
\hline
$M_n$ & 3 & 10 & 35 & 126 & 462 & 1716 & 6435 & 24310 & 92378 & 352716 \\
\hline
Permanent-Based & 36s & 3m & 16m & 1h30 & 11h & 3d & \multicolumn{3}{c}{\it not possible}&\\
\hline
\SLOS & 14s & 15s & 29s & 73s & 2m & 5m & 22m & 1h45m & 10h & 2d8h \\
\hline
\end{tabular}
    \caption{In this table, we compare time of QML algorithm to perform 150 optimization iterations with both Permanent-Based and \SLOS algorithms. This application is the ideal use case of \SLOS since we are interested into the full output states exact distribution.}
    \label{tab:perf_qml}
\end{table}

\begin{figure}[t]
\centering
    \includegraphics[width=10cm]{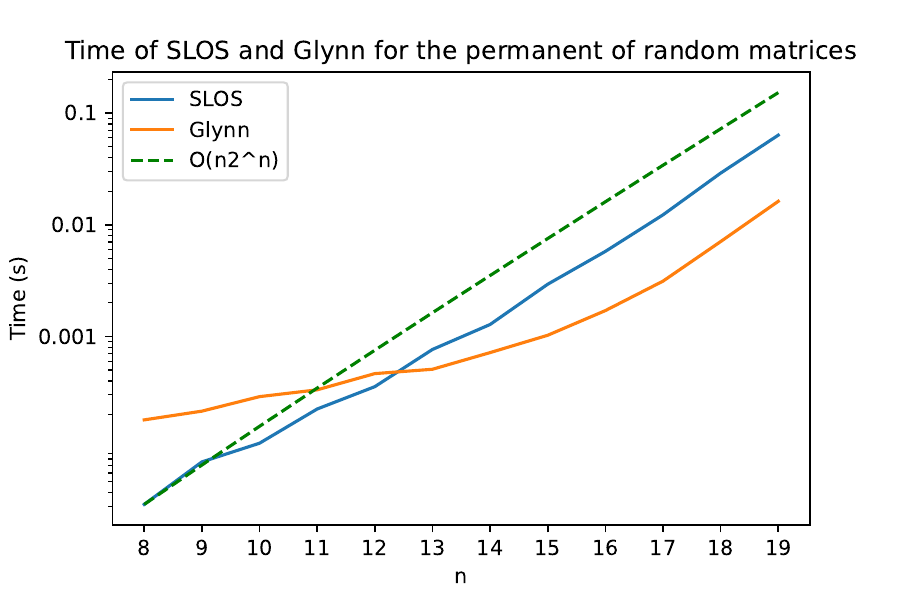}
    \caption{\label{fig:glynnslos}Comparative performance of Glynn and \SLOS for calculation of a single $n\times n$ permanent 128-bit complex numbers. The green curve is the $n2^n$ curve renormalised with the first point of the \SLOS curve. Benchmark computes 100 permanents in a row and outputs the mean time. The benchmark is ran on a Intel Core i7-10510U with 16GB of memory.
      }
   
\end{figure}

\subsection{Benchmarking \SLOSgen for One Output}
\label{sec:performance_single}

We compare in \cref{fig:glynnslos} the speed of our algorithm to
compute a single permanent in the worst possible situation, i.e. when the
output is $\ket{1,....,1}$, with a traditional permanent calculation
algorithm. We selected the algorithm from Glynn
\cite{glynn2010permanent} as \cite{heurtel_perceval_2022} shows that
for up to 19 modes, this algorithm is practically one of
the most efficient.


The curves does not show the precompilation time needed for
  \SLOS nor the allocation of the Fock states.  We can see that
  the practical time for \SLOS is better for small cases, and that, as
  predicted by the complexity analysis, the growth is very close to the
  $n2^n$ curve. This is even true for small instances.


The code for Glynn's algorithm developed in \emph{Perceval}
  \cite{heurtel_perceval_2022} makes heavy use of hardware
  optimizations, and shows a speedup compared to \SLOS when $n>12$.
  The speedup has a multiplicative factor between 2 and 4 depending on
  the machine. On Figure \ref{fig:glynnslos}, the factor is around 4
  for large values of $n$. Indeed, as the data in Glynn's algorithm is
  local, one can rely on efficient libraries to sumprod the terms,
  improving the time of the computations. Due to the structure of the
  algorithm (because of the heavy use of pre-computed data), \SLOS
  cannot make use of these optimizations.
  One has however to note that without these hardware-specific
  optimizations, there is no particular speedup for Glynn's algorithm
  compared to \SLOS.

\begin{figure}[t]
    \centering
    \includegraphics[width=10cm]{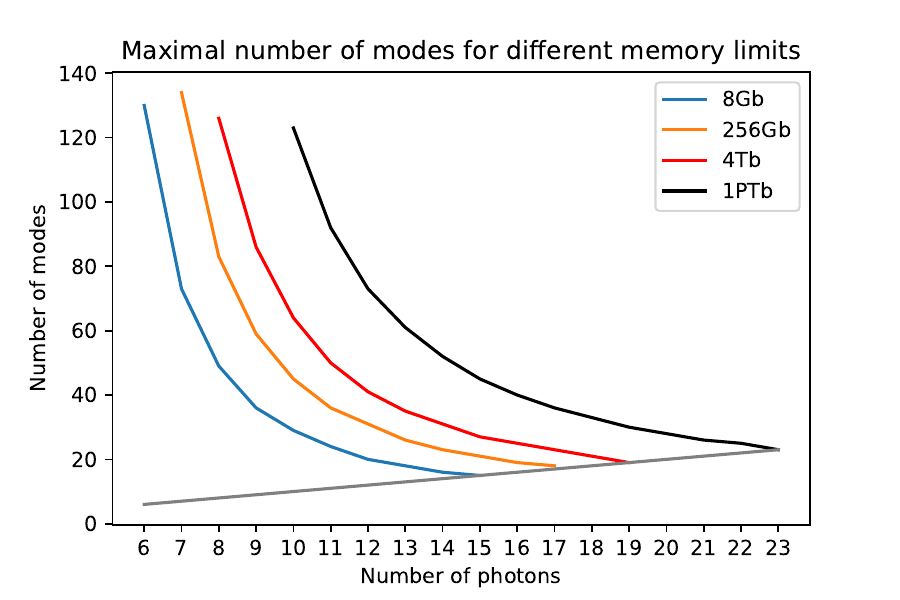}
    \caption{\label{fig:slosmemorylimits}This graph shows four regions for $(m,n)$ combinations. In blue - $m\ge n$ with configurations that can run on a personal laptop (up to 8Gb memory). Curves in orange/red/black are the respective limits of 256Gb/4Tb/1Pb memory. For instance, strong simulation with \SLOS of 24 photons on a 24 modes circuit would require $1500Tb$ of processing memory!}    
\end{figure}

\subsection{Memory Usage}
\label{sec:memory}

Finally, since \SLOSgen is making intensive use of memory, we have computed in \cref{fig:slosmemorylimits} the practical limitations of strong simulation. The graphics vertical axis is the number of modes $m$, and the horizontal axis the number of photons $n$. We are only interested in $m\ge n$, and we have drawn boundaries of three typical workstations. In blue, computer with up to 8Gb of memory - typically any modern laptop. In orange, the limit of 256 Gb memory - typically a large compute node. Any $(m,n)$ point below the orange curve will fit in 256 Gb memory. The red curve represents memory need up-to 4 Terabytes memory representing a very large HPC node. The black curve represents a potential super-computer with up to 1 Petabyte of memory. As of today technology this boundary can be considered as an area unreachable for strong simulation, and so for a full description of linear optical processes.



\section{Conclusion}
\label{sec:conclusion}

In this paper, we presented a versatile framework for the simulation of linear optical circuits, with a trade-off between time and memory usage. An efficient implementation is provided and we discuss how it outperforms the permanent-based algorithms. It is an open problem to determine to what extend the memory usage of \SLOS can be improved, or to what extend the time complexity of the permanent-based method can be improved with more memory.
As a future work, we plan to incorporate noise models, and validate the simulations against physical hardware.


\section*{Acknowledgement}

The authors would like to thank Rawad Mezher and Timothée Goubault de Brugière for helpful discussions and references, and the two reviewers for their very valuable comments, that have greatly improved the quality of the paper.
This work is supported by the PEPR integrated project EPiQ ANR-22-PETQ-0007 part of Plan France 2030 and by the CIFRE 2022/0081.
It is also supported by the French National Research Agency (ANR) under the research projects
SoftQPro ANR-17-CE25-0009-02, by the STIC-AmSud project Qapla’ 21-STIC-10, and
by the European project HPCQS.


\bibliographystyle{linksen}

\bibliography{reference}

\end{document}